\documentclass[authoryear,preprint,12pt]{elsarticle}

\usepackage{amssymb}
\usepackage{amsmath}
\usepackage{xcolor}

\newcommand{\edit}[1]{#1}
\usepackage[
    colorlinks=true,
    linkcolor=blue,
    citecolor=blue,
    urlcolor=blue
]{hyperref}
\hypersetup{
    pdfauthor={Jeongdong Kim, Jonggeol Na, Sungho Shin}
}
\usepackage[nameinlink]{cleveref}
\usepackage{booktabs}
\crefname{equation}{}{}
\Crefname{equation}{Equation}{Equations}
\crefname{table}{Table}{Tables}
\Crefname{table}{Table}{Tables}
\crefname{figure}{Figure}{Figures}
\Crefname{figure}{Figure}{Figures}
\crefname{section}{Section}{Sections}
\Crefname{section}{Section}{Sections}

\journal{Applied Energy}

\begin{document}

\begin{frontmatter}

\title{\edit{Reliability Value of Long-Duration Energy Storage against Extreme Events in High-Renewable Grids: A Full-Year AC-OPF Assessment}} %% Article title

\author[1,2]{Jeongdong Kim}
\author[3,4,5]{Jonggeol Na\corref{cor1}}
\ead{jgna@ewha.ac.kr}
\author[1]{Sungho Shin\corref{cor1}}
\ead{sushin@mit.edu}
\cortext[cor1]{Corresponding authors}

\affiliation[1]{
    organization={Department of Chemical Engineering, Massachusetts Institute of Technology},
    addressline={77 Massachusetts Ave},
    city={Cambridge},
    postcode={02139},
    state={Massachusetts},
    country={United States}
}
\affiliation[2]{
    organization={Department of Chemical and Biomolecular Engineering, Yonsei University},
    addressline={50 Yonsei-ro},
    city={Seoul},
    postcode={03722},
    state={},
    country={Republic of Korea}
}
\affiliation[3]{
    organization={Department of Chemical Engineering and Materials Science, Ewha Womans University},
    addressline={52 Ewhayeodae-gil},
    city={Seoul},
    postcode={03760},
    state={},
    country={Republic of Korea}
}
\affiliation[4]{
    organization={Graduate Program in System Health Science and Engineering, Ewha Womans University},
    addressline={52 Ewhayeodae-gil},
    city={Seoul},
    postcode={03760},
    state={},
    country={Republic of Korea}
}
\affiliation[5]{
    organization={Institute for Multiscale Matter and Systems (IMMS), Ewha Womans University},
    addressline={52 Ewhayeodae-gil},
    city={Seoul},
    postcode={03760},
    state={},
    country={Republic of Korea}
}

%% Abstract
\begin{abstract}
\edit{Long-duration energy storage (LDES) can mitigate prolonged renewable--load imbalances during Dunkelflaute events, but existing studies rely on zonal or linearized DC network models and inadequately analyze the operational feasibility of the grid across a wide range of full-year renewable, load, and contingency scenarios. To address this gap, this paper introduces a multi-period alternating-current optimal power flow (AC-OPF) formulation that captures the full nonlinear network physics and assesses the reliability value of LDES in high-renewable grids. To evaluate scarcity events spanning multiple days to weeks, full-year operation is modeled as an 8,760-h load-shedding minimization over scenarios sampled from a Gaussian copula model, fitted to 2010--2025 historical wind and bus-level load data, that represents both typical variability and tail events. On a synthetic 200-bus Illinois transmission network, a hybrid fleet of battery energy storage (BESS) and LDES with 50 MW total power reduces annual load shedding by 83.0\% on average relative to the base network, versus 68.5\% for short-duration BESS alone at equal power. To further account for unexpected line outages throughout the year, the formulation is extended to a multi-day security-constrained AC-OPF. Under $N-1$ contingencies, no feasible operating solution is obtained for the base network, whereas the LDES-equipped network remains feasible in all considered cases, thereby saving the cost of additional generation and transmission capacity. During the contingency period, LDES acts as a backup power supply, requiring only 5.5\% more generation on average than the no-contingency base case.}
\end{abstract}

% \begin{highlights}
% \item \edit{Multi-period AC-OPF, representing the full network physics and storage constraints, is augmented with statistical sampling studies to characterize tail events and quantify the reliability value of LDES.}
% \item \edit{Duration itself carries reliability value: at equal storage power capacity, the hybrid BESS--LDES fleet cuts annual load shedding by 83.0\% on average versus 68.5\% for short-duration BESS alone.}
% \item \edit{$N-1$ security conventionally requires idle backup capacity; at equal storage power, short-duration BESS fails every contingency case, whereas the BESS--LDES fleet sustains all of them with only 5.5\% more generation.}
% \end{highlights}

%% Keywords

\end{frontmatter}

\section{Introduction}
\label{Intro}
% Necessity of energy storage system in high renewable penetration
\edit{During Dunkelflaute events, extended periods of low wind and solar availability, grids with high renewable penetration can face supply shortages lasting from several days to weeks \citep{denholm2021challenges,raynaud2018energy,van2024temporally}. Long-duration energy storage (LDES) can bridge such prolonged scarcity by sustaining discharge over multi-day horizons. In current practice, however, deployed storage consists mostly of battery energy storage systems (BESS) with short durations of 2--8~h, favored for their high round-trip efficiency of around 0.9 \citep{viswanathan20222022}. Such short durations cannot mitigate energy mismatches that persist over multiple days. On the demand side, rapid growth in data-center deployment further reshapes load profiles and increases the risk of unexpected network stress through large and variable AI workloads \citep{li2024unseen,lin2024exploding}.}

% Intro of LDES and current research question
Prior studies have justified the system-level operational benefits and economic feasibility of LDES across diverse technical and cost parameter settings \citep{sepulveda2021design,jenkins2017enhanced,zhang2020benefit}. \edit{Evaluating the reliability value of LDES on an existing grid, however, requires more than system-level energy accounting. Network reliability is determined by whether the available generation, transmission, and storage resources can serve load across diverse scenarios while respecting the physical governing equations and the operational limits of the network. Reactive power and voltage control are integral to this assessment: power transfers can be limited by voltage constraints as much as by thermal limits, and inverter-based storage contributes reactive power support in addition to active power, effects that are visible only under the full AC power flow physics. Assessing this adequately calls for optimization problems constrained by the AC power flow equations, combined with realistic models of uncertainty and dynamic load variation, given the existing generation, transmission, and storage capacities and the storage duration.} Moreover, energy storage systems are expected to support reliable power supply during unpredictable events such as transmission-line outages. \edit{Ensuring $N-1$ security conventionally requires additional generation or transmission capacity that remains largely idle outside contingency events; storage that is actively used in normal operation can potentially substitute for such capacity.} These network-level perspectives are timely, as U.S. regulatory practice emphasizes resource accreditation for LDES with clear justification of its operational characteristics under realistic network conditions \citep{cpuc}.

% Brief motivation and contribution of this study
\edit{These considerations, the multi-day scarcity that LDES must bridge and the physical and security requirements that grid operation must satisfy, lead to the central question addressed in this study:}
\begin{quote}
\edit{\emph{How can the reliability value of LDES in a high-renewable grid be realistically quantified, respecting the nonlinear network physics and operational limits, under full-year renewable and load uncertainty and $N-1$ line contingencies?}}
\end{quote}
\paragraph{Our approach}
\edit{We answer this question through two complementary analyses.} For the first, we formulate a full-year \edit{alternating-current optimal power flow (AC-OPF)} problem with load-shedding minimization (\cref{subsec:full_year_opf}). A Gaussian copula method is employed to construct full-year renewable and load uncertainty scenarios (\cref{subsec:Uncertainty-modeling}). By repeatedly solving the AC-OPF across the scenario set, the resulting solution set is translated into network-feasibility metrics to quantify the impact of LDES technical parameters on transmission-network operation (\cref{subsec:full_year_result}). \edit{Using the identified LDES parameters, further optimization with generation minimization quantifies the renewable curtailment reduction and the resulting thermal generation savings enabled by the storage (\cref{subsec:generation_analysis}).} For the second, \edit{we assess the value of LDES in substituting the capacity that would otherwise be required to guarantee $N-1$ security; the $N-1$ criterion of established reliability planning requires the grid to remain operable after the loss of any single component. The instrument for this assessment is} a multi-day SCOPF that jointly considers the contingency period and the restoration period (\cref{subsec:SC-AC-OPF}). \edit{ Considering 24-hour line outages with occurrence times spanning the full year covers contingencies striking during Dunkelflaute-driven scarcity periods, when the grid is most stressed and the seasonal LDES energy state governs the available backup power.} This formulation enables us to evaluate whether LDES can support corrective redispatch during $N-1$ line outages while preserving sufficient stored energy to return to the nominal full-year operational trajectory following the contingency (\cref{subsec:contingecy_analysis}). \edit{Solving such multi-period AC-OPF and SC-AC-OPF problems at the full-year scale was previously considered computationally intractable; with the recent emergence of GPU-based nonlinear optimization solvers, these problems can now be solved efficiently \citep{johnson2025examodelspower,shin2024accelerating}.}

% Overveiw of LDES-grid study
\paragraph{Related works}
\edit{Early studies of LDES center on modeling its techno-economic characteristics and optimizing its deployment at the system level. Because LDES installation requires additional investment cost, these studies identify the requirements for cost-effective deployment, such as power and energy capacity, efficiency, storage duration, and capital cost, while evaluating system-level benefits such as curtailment reduction and seasonal energy shifting.} For example, \edit{\citet{sepulveda2021design} cover} the LDES design space with diverse electrification scenarios of the network and \edit{employ} a capacity expansion model-based optimization \citep{jenkins2017enhanced}. The results \edit{demonstrate} that capacity cost and discharge efficiency have the greatest impact on power generation cost savings, and the core finding is the upper bound on LDES energy capacity cost for full network electrification under the target site's weather and demand conditions.

For narrowing to the operational level, \edit{\citet{zhang2020benefit}} simulate a full-year unit commitment problem to assess the economic benefits of LDES. Under a 2050 Western Interconnection scenario with 85\% renewable penetration, a 2,000 MW LDES system with an 8 h duration reduces fuel consumption as well as generator start-up and shutdown costs through both diurnal and seasonal operation. In addition, renewable curtailment is reduced from 2.4 TWh to 2.2 TWh. These results demonstrate the benefits of LDES, including improved generator commitment decisions and reduced renewable curtailment.

\edit{Using a macro-scale energy model over 39 weather years, \citet{dowling2020role} demonstrate that in least-cost wind--solar systems, batteries provide intra-day balancing while LDES provides seasonal and even multi-year storage. Most directly related to scarcity events, \citet{kittel2026longduration} quantify the European LDES capacity required to cope with Dunkelflaute events using power-sector modeling over 35 weather years, and demonstrate that the most extreme scarcity events, rather than typical years, determine LDES sizing and operation.}

More recent work has begun to move beyond planning and incorporate physical constraints of the power network. \edit{\citet{chu2025long} identify the optimal siting of LDES considering} transmission line capacity. Their results \edit{demonstrate} that transmission capacity strongly affects optimal LDES siting, indicating that LDES value depends on both spatial supply--demand patterns and network structure. Additionally, \edit{\citet{piansky2024long}} formulate a full-year linearized network power flow model for LDES siting and sizing. Their study considers network failure events and optimizes hourly operation.

% Brief summary of OPF
In parallel, hourly operation of the transmission network can be formulated as an optimal power flow problem. \edit{The linearized DC-OPF variant is widely used for its tractability, but it neglects reactive power and voltage constraints altogether, and its solutions are never feasible with respect to the AC power flow equations \citep{baker2021never}; network feasibility analysis therefore calls for the AC-OPF formulation,} which explicitly enforces nonlinear power balance, voltage limits, line-flow limits, and device-capacity constraints \citep{gayme2012optimal,gabash2012active}. \edit{Because storage couples operating decisions across time and scarcity events span multiple days, multi-period AC-OPF, which incorporates the inter-temporal coupling constraints of energy storage, enables a quantitative assessment of the feasibility value of energy storage in network operation.} Unlike conventional generation-cost minimization, the problem can be reformulated to minimize load-shedding costs \citep{majumdar1996interruptible,xu2001optimal,hazra2007congestion} or corrective re--dispatch costs \citep{martins2008redispatch,thukaram2008optimal}, thereby quantifying the ability of LDES to mitigate the network's operational constraint violations. \edit{In reliability planning, adequacy is conventionally quantified with probabilistic indices such as the loss-of-load expectation, including the widely used one-day-in-ten-years criterion, and the expected unserved energy \citep{billinton1996reliability}; in this study, we instead assess reliability through the tail of the load-shedding distribution, using the 99th-percentile annual apparent load shedding across scenarios, which directly captures performance under extreme events. Finally, because reliable operation must also withstand unexpected component outages,} security-constrained OPF (SCOPF) further extends this formulation by enforcing feasibility under contingency scenarios, including $N-1$ line outages and other stress events \citep{stott2005security,capitanescu2011state}. 

% Current limitation
Despite extensive prior work, \edit{the reliability value of LDES} remains insufficiently investigated. The aforementioned studies \citep{sepulveda2021design,jenkins2017enhanced,zhang2020benefit,dowling2020role,kittel2026longduration} provide important insights into the system-level operational and economic value of LDES, but \edit{represent the transmission network either not at all or in simplified form (linearized power flow or line-capacity limits alone), which cannot verify whether an LDES-equipped grid satisfies the nonlinear AC power flow equations, voltage limits, and line capacities}. Conversely, AC-OPF studies provide quantitative assessments of network feasibility under nonlinear physical constraints, but are typically conducted for short-duration storage (e.g., 4 hours) \citep{marley2016multi}, over limited operational horizons (e.g., 24 hours) \citep{soares2017active,alizadeh2022envisioning}. Because \edit{different LDES technologies span storage durations of 10--100 hours or longer}, network-level operation should be evaluated over full-year horizons to capture \edit{long-term and seasonal storage operation}. Moreover, current regulatory practices specify operational criteria for LDES deployment in networks. For instance, the California Public Utilities Commission (CPUC) currently categorizes an LDES as an energy resource capable of lasting 8 hours maximum discharging \citep{cpuc}. Moreover, the CPUC requires the operational characteristics of LDES to be evaluated under different operating conditions, including both base-case and contingency conditions. \edit{Taken together, the literature and emerging regulatory rules highlight the need to quantify the reliability value of LDES at the transmission level, under full-year uncertainty including Dunkelflaute-driven scarcity and under $N-1$ contingency events; the value of storage depends not only on system-level energy shifting, but also on whether the existing grid can reliably deliver it under realistic operating conditions.}

\edit{The most relevant prior works to this study are the system-level analyses of \citet{dowling2020role} and \citet{kittel2026longduration}, which establish the seasonal balancing role of LDES and its sizing against Dunkelflaute events without network physics; the network-aware LDES studies of \citet{chu2025long} and \citet{piansky2024long}, which represent the transmission network in simplified form; and the multi-period AC-OPF studies of \citet{marley2016multi} and \citet{soares2017active}, which consider short-duration storage over short horizons.}

\paragraph{Contributions}
\edit{The main contribution of this study is a multi-period AC-OPF-based characterization of the reliability value of LDES in high-renewable grids. Relative to the works above, our contributions are as follows:}
\begin{itemize}
\item \edit{We formulate a full-year, 8,760-h multi-period AC-OPF with load-shedding minimization that couples the inter-temporal constraints of BESS and LDES with the nonlinear AC network physics, and demonstrate that these problems, previously considered computationally intractable, can be solved at the full-year scale with GPU-based nonlinear optimization solvers.}
\item \edit{We introduce the \emph{seasonally anchored} SC-AC-OPF to assess how much of the additional capacity buildout required for $N-1$ security can be avoided by extending the storage duration; the formulation couples the security assessment with the seasonal LDES energy state.}
\item \edit{We develop reliability metrics and diagnostics, the 99th-percentile annual apparent load shedding and constraint-binding-hour attribution, and use them to quantify the reliability value of LDES across its technical parameters, identifying configurations that eliminate load shedding and sustain secure operation under all considered contingencies, without additional generation or transmission expansion.}
\end{itemize}
\edit{To the best of our knowledge, this is the first assessment of LDES against nonlinear AC network physics over a full-year horizon under renewable and load uncertainty and $N-1$ contingencies.}

\edit{The rest of this article is organized as follows. \Cref{method} presents the problem formulations, uncertainty modeling, and \edit{case study setup}, \Cref{sec:results} reports the numerical results, and \Cref{sec:conclusion} concludes this article.}

\section{Methodology}
\label{method}
\edit{An overview of the methodology is as follows. The reliability value of LDES is assessed under two operating cases: a nominal case, in which full-year operation under renewable and load uncertainty is evaluated through a load-shedding minimization AC-OPF, and a contingency case, in which secure operation under $N-1$ line outages is evaluated through a multi-day security-constrained AC-OPF (SC-AC-OPF), quantifying the capacity buildout that LDES avoids.} First, the full-year AC-OPF is repeatedly solved under load and renewable generation scenarios augmented using a Gaussian copula method \citep{nelsen2006introduction}. The resulting optimal solution sets are translated into the 99th-percentile annual load-shedding metric. A parametric sweep over LDES capacity, round-trip efficiency, and duration is then conducted to reveal their impact on network feasibility and identify the parameter settings that achieve full-year operation without load shedding \edit{across the scenario set. For these zero-shedding configurations, an additional generation-minimization solve quantifies the renewable curtailment absorbed by the storage (\cref{subsec:generation_analysis})}. Using the identified LDES parameters, the SC-AC-OPF is iteratively solved under $N-1$ line contingencies by shifting the contingency occurrence time through the year, thereby covering diverse operating conditions including peak load. The full-year AC-OPF and multi-day SC-AC-OPF formulations are described in \cref{subsec:full_year_opf} and \cref{subsec:SC-AC-OPF}, respectively, while the uncertainty modeling is presented in \cref{subsec:Uncertainty-modeling}. \edit{The case study setup}, including the network data, NLP solver, convergence tolerances, and computational resources, \edit{is} provided in \cref{subsec:implementation}.

\subsection{Full-year AC-OPF with LDES}
\label{subsec:full_year_opf}
\subsubsection{Multi-period AC-OPF}
\label{subsubsec:ac_opf}
We use a multi-period AC-OPF formulation with load-shedding minimization to assess network feasibility under renewable penetration and LDES operation over a full-year horizon $\mathcal{T}=\{1,\ldots,8760\}$. The formulation extends the standard AC-OPF model \citep{momoh2002economic,ge1999optimal} by incorporating \edit{renewable generators, BESS, and LDES devices, together with a relaxation of the power balance constraints}. The sets $\mathcal{N}$, $\mathcal{L}$, and $\mathcal{T}$ denote buses, transmission lines, and time periods, respectively. The sets $\mathcal{D}$, $\mathcal{G}$, $\mathcal{R}$, and $\mathcal{S}$ denote loads, thermal generators, renewable generators, and energy-storage systems (ESS). The subscript $n$ indicates the subset of each component connected to bus $n$. The subsets $\mathcal{L}^{\mathrm{from}}_n$ and $\mathcal{L}^{\mathrm{to}}_n$ denote the sets of lines directed from and to bus $n$, respectively.

Given bus $n$, the active and reactive power balance at time $t$ is:
{\small
\begin{subequations}
\begin{align}
& \sum_{l \in \mathcal{L}^{\mathrm{to}}_n} p_{lt}
- \sum_{l \in \mathcal{L}^{\mathrm{from}}_n} p_{lt}
+ \sum_{g \in \mathcal{G}_n} p^G_{gt}
+ \sum_{r \in \mathcal{R}_n} p^R_{rt} \notag \\
& \quad = \sum_{s \in \mathcal{S}_n} p^{ES}_{st}
+ \sum_{d \in \mathcal{D}_n}
\left(P^D_{dt}-P^{\mathrm{LS}}_{dt}\right)
\quad \forall n \in \mathcal{N},\ t \in \mathcal{T},
\label{eq:p_balance} \\
& \sum_{l \in \mathcal{L}^{\mathrm{to}}_n} q_{lt}
- \sum_{l \in \mathcal{L}^{\mathrm{from}}_n} q_{lt}
+ \sum_{g \in \mathcal{G}_n} q^G_{gt}
+ \sum_{r \in \mathcal{R}_n} q^R_{rt} \notag \\
& \quad = \sum_{s \in \mathcal{S}_n} q^{ES}_{st}
+ \sum_{d \in \mathcal{D}_n}
\left(Q^D_{dt}-Q^{\mathrm{LS}}_{dt}\right)
\quad \forall n \in \mathcal{N},\ t \in \mathcal{T},
\label{eq:q_balance} \\
& 0 \le P^{\mathrm{LS}}_{dt} \le P^D_{dt},
\quad \forall d \in \mathcal{D},\ t \in \mathcal{T}
\label{eq:acopf_ls_bound}
\end{align}
\end{subequations}}\noindent
\edit{Here, $p_{lt}$ and $q_{lt}$ denote the active and reactive power flows on line $l$ at time $t$; $p^G_{gt}$ and $q^G_{gt}$ denote the active and reactive power outputs of thermal generator $g$; $p^R_{rt}$ and $q^R_{rt}$ denote those of renewable generator $r$; and $p^{ES}_{st}$ and $q^{ES}_{st}$ denote the active and reactive powers injected from the bus into energy storage system $s$. The parameters $P^D_{dt}$ and $Q^D_{dt}$ denote the active and reactive loads of demand $d$, while $P^{\mathrm{LS}}_{dt}$ and $Q^{\mathrm{LS}}_{dt}$ denote the corresponding active and reactive load-shedding variables.}

Following ISO network-modeling guidelines \citep{caisoBPMFNM,huang2002voltage}, reactive load shedding in \cref{eq:q_balance} is modeled as proportional to active load shedding under a constant-power-factor assumption:
\begin{equation}
Q^{\mathrm{LS}}_{dt}
= Q^D_{dt}\frac{P^{\mathrm{LS}}_{dt}}{P^D_{dt}} ,
\quad \forall d \in \mathcal{D},\ t \in \mathcal{T}
\label{eq:constant_pf_ls}
\end{equation}

For each line $l$ at time $t$, the AC line flow is \edit{given} as follows:
{\small
\begin{subequations}
\begin{align}
& p_{lt} = v_{mt}v_{nt}
\left[
G_l\cos(\theta_{nt}-\theta_{mt})
+B_l\sin(\theta_{nt}-\theta_{mt})
\right] \label{eq:p_flow}\\
& \quad \forall l=(m,n)\in\mathcal{L},\ t\in\mathcal{T},\notag \\
& q_{lt}= v_{mt}v_{nt}
\left[
G_l\sin(\theta_{nt}-\theta_{mt})
-B_l\cos(\theta_{nt}-\theta_{mt})
\right] \label{eq:q_flow}\\
& \quad \forall l=(m,n)\in\mathcal{L},\ t\in\mathcal{T},\notag \\
& V^{\mathrm{min}}_{n} \leq v_{nt} \leq V^{\mathrm{max}}_{n} \quad \forall n\in\mathcal{N},\ t\in\mathcal{T}, \label{eq:v_range}\\
& \Theta^{\mathrm{min}}_{mn} \leq \theta_{mt}-\theta_{nt} \leq \Theta^{\mathrm{max}}_{mn} \quad \forall (n,m)\in\mathcal{L},\ t\in\mathcal{T} \label{eq:phase_range}
\end{align}
\end{subequations}}\noindent where $m$, $n$ \edit{denote} the connected bus nodes on line $l$, $G_l$ and $B_l$ denote the conductance and susceptance of line $l$, respectively, \edit{$v_{nt}$ is the voltage magnitude} at bus $n$ at time $t$, bounded by $V^{\mathrm{min}}_{n}$ and $V^{\mathrm{max}}_{n}$, and $\theta_{nt}$ is \edit{the phase angle at bus $n$, with the angle difference between connected buses bounded by} $\Theta^{\mathrm{min}}_{mn}$ and $\Theta^{\mathrm{max}}_{mn}$.

The apparent power flow on each line is constrained by the thermal capacity $S_l^{\max}$:
\begin{equation}
\sqrt{p_{lt}^{2}+q_{lt}^{2}}
\le S_l^{\max}\quad \forall l \in \mathcal{L},\ t \in \mathcal{T}
\label{eq:line_limit} 
\end{equation}

This study considers two types of generators, namely thermal \edit{generators} $g \in \mathcal{G}$ and renewable generators $r \in \mathcal{R}$\edit{,} as follows:
{\small
\begin{subequations}
\begin{align}
& P_g^{\min} \le p^G_{gt} \le P_g^{\max}
\quad \forall g \in \mathcal{G},\ t \in \mathcal{T},
\label{eq:gen_p_bound} \\
& \left|p^G_{gt}-p^G_{g,t-1}\right| \leq P^{\mathrm{ramp}}_g \quad \forall g \in \mathcal{G},\ t \in \mathcal{T}\setminus\{1\}, 
\label{eq:gen_p_ramp} \\
& Q_g^{\min} \le q^G_{gt} \le Q_g^{\max}
\quad \forall g \in \mathcal{G},\ t \in \mathcal{T},
\label{eq:gen_q_bound} \\
& 0 \le p^R_{rt} \le \min\left\{P_r^{\max}, \overline{P}^{R}_{rt}\right\}
\quad \forall r \in \mathcal{R},\ t \in \mathcal{T},
\label{eq:renewable_p_capacity_bound} \\
& Q_r^{R,\min} \le q^R_{rt} \le Q_r^{\max}
\quad \forall r \in \mathcal{R},\ t \in \mathcal{T},
\label{eq:renewable_q_bound} \\
& \left|p^R_{rt}-p^R_{r,t-1}\right| \leq P^{\mathrm{ramp}}_r \quad \forall r \in \mathcal{R},\ t \in \mathcal{T}\setminus\{1\} 
\label{eq:renew_p_ramp}
\end{align}
\end{subequations}}\noindent where $(P_g^{\min},P_g^{\max})$ and $(Q_g^{\min},Q_g^{\max})$ denote the active- and reactive-power limits of thermal generator $g$, $P_r^{\max}$ and $(Q_r^{R,\min},Q_r^{\max})$ denote the active-power capacity and reactive-power limits of renewable generator $r$, $\overline{P}^{R}_{rt}$ denotes the time-dependent renewable power availability, and $P_g^{\mathrm{ramp}}$ and $P_r^{\mathrm{ramp}}$ denote the ramping limits of thermal and renewable generators, respectively. 

\subsubsection{Energy storage modeling}
\label{subsubsec:ess_model}
\edit{This study abstracts away the technology-specific characteristics of LDES: candidate technologies, such as pumped hydro storage, compressed-air energy storage, hydrogen-based storage, thermal energy storage, and flow batteries \citep{sepulveda2021design}, are represented by a common storage model parameterized by round-trip efficiency, storage duration, and power capacity, with the parameter ranges considered in this study (RTE of 0.3--0.8 and durations of 20--100 h) spanning these technologies.}

Given bus $n$, we assume that a BESS and \edit{an} LDES are co-located at each ESS $s \in \mathcal{S}_n$. Accordingly, \edit{the} power injected from \edit{the} bus to \edit{the} storage, $(p^{\mathrm{ES}}_{st}, q^{\mathrm{ES}}_{st})$, is defined as the sum of corresponding BESS and LDES power components:
{\small
\begin{subequations}
\begin{align}
p^{\mathrm{ES}}_{st}
&= p^{\mathrm{B}}_{st} + p^{\mathrm{LD}}_{st} 
\quad \forall s \in \mathcal{S},\ t \in \mathcal{T},
\label{eq:ess_active_injection} \\
q^{\mathrm{ES}}_{st}
&= q^{\mathrm{B}}_{st} + q^{\mathrm{LD}}_{st}
\quad \forall s \in \mathcal{S},\ t \in \mathcal{T}
\label{eq:ess_reactive_injection}
\end{align}
\end{subequations}}\noindent where ($p^{\mathrm{B}}_{st}$, $q^{\mathrm{B}}_{st}$) and ($p^{\mathrm{LD}}_{st}$, $q^{\mathrm{LD}}_{st}$) denote the powers injected from the bus to the inverter-based power conversion system of BESS and LDES. 

For each storage type $h \in \mathcal{H}=\{\mathrm{B},\mathrm{LD}\}$, the stored energy $E^h_{st}$ is updated with charging and discharging efficiencies, and the corresponding operating limits are imposed as follows:
{\small
\begin{subequations}
\begin{align}
& E^{h}_{st}
= E^{h}_{s,t-1}
+ \eta^{\mathrm{CH},h}_{s}p^{\mathrm{CH},h}_{st}
- \frac{p^{\mathrm{DH},h}_{st}}{\eta^{\mathrm{DH},h}_{s}}
\quad \forall \edit{s \in \mathcal{S},\ }t \in \mathcal{T}\setminus\{1\}, \label{eq:storage_energy_update} \\
& E^{h}_{s1}
= E^{h}_{s,init}
+ \eta^{\mathrm{CH},h}_{s}p^{\mathrm{CH},h}_{s1}
- \frac{p^{\mathrm{DH},h}_{s1}}{\eta^{\mathrm{DH},h}_{s}} \quad \edit{\forall s \in \mathcal{S}}, \label{eq:init_store_update} \\
& 0 \le p^{\mathrm{CH},h}_{st}
\le P^{h,\max}_{s} \quad \forall \edit{s \in \mathcal{S},\ }t \in \mathcal{T},
\label{eq:storage_charge_bound} \\
& 0 \le p^{\mathrm{DH},h}_{st}
\le P^{h,\max}_{s} \quad \forall \edit{s \in \mathcal{S},\ }t \in \mathcal{T},
\label{eq:storage_discharge_bound} \\
& E^{h,\min}_{s}
\le E^{h}_{st}
\le E^{h,\max}_{s} \quad \forall \edit{s \in \mathcal{S},\ }t \in \mathcal{T},
\label{eq:storage_energy_bound} \\
& E^{h,\max}_{s}
= H^{h}_{s} P^{h,\max}_{s} \quad \edit{\forall s \in \mathcal{S}}
\label{eq:storage_energy_capacity}
\end{align}
\end{subequations}}\noindent where $\mathcal{H}=\{\mathrm{B},\mathrm{LD}\}$ denotes the storage type, corresponding to BESS and LDES, respectively, $p^{\mathrm{CH},h}_{st}$ and $p^{\mathrm{DH},h}_{st}$ denote the charging and discharging powers of storage type $h$, and $\eta^{\mathrm{CH},h}_{s}$ and $\eta^{\mathrm{DH},h}_{s}$ denote the charging and discharging efficiencies. \edit{The parameter $E^{h}_{s,init}$ denotes the initial stored energy, and $E^{h,\min}_{s}$ and $E^{h,\max}_{s}$ denote the minimum and maximum energy bounds.} The charging and discharging powers are bounded by $P^{h,\max}_{s}$, and the maximum energy capacity is \edit{given} by the storage duration $H^h_s$ \edit{through \cref{eq:storage_energy_capacity}}.

We adopt the inverter-based grid-interface model from \edit{\citet{geth2020flexible}} to represent the bus-storage power exchange. For storage type $h$ installed at site $s$ and connected to bus $n$, the grid-interface model is formulated for all $t \in \mathcal{T}$ as follows:
{\small
\begin{subequations}
\begin{align}
&p^{h}_{st} = p^{\mathrm{CH},h}_{st} - p^{\mathrm{DH},h}_{st} + R^{h}_{s}(I^{h}_{st})^2,
\label{eq:active_injection} \\
&q^{h}_{st} = q^{\mathrm{int},h}_{st} + X^{h}_{s}(I^{h}_{st})^2,
\label{eq:reactive_injection} \\
&\left(p^{h}_{st}\right)^2
+\left(q^{h}_{st}\right)^2
=v_{nt}^{2}(I^{h}_{st})^2,
\label{eq:current_relation} \\
&\left(p^{h}_{st}\right)^2
+\left(q^{h}_{st}\right)^2
\le \left(S^{h,\max}_{s}\right)^2,
\label{eq:converter_flow_limit} \\
& S^{h,\max}_{s} = 1.05 P^{h,\max}_{s},
\label{eq:convert_max_rate} \\
& -0.33 P^{h,\max}_{s} \le q^{h}_{st}
\le 0.33 P^{h,\max}_{s}
\label{eq:convert_reactive_range}
\end{align}
\end{subequations}}\noindent where $I^{h}_{st}$ is the inverter current, $q^{\mathrm{int},h}_{st}$ is the internal reactive-power variable, $v_{nt}$ denotes the voltage magnitude at the connected bus $n$, and $R^h_{s}$ and $X^h_{s}$ are the inverter resistance and reactance parameters, which are set to 0.002 p.u. and 0.08 p.u., respectively \citep{karlson2012wind}.

\Cref{eq:active_injection,eq:reactive_injection} define the power injections from the connected bus to storage type $h$, while \cref{eq:current_relation,eq:converter_flow_limit} couple these injections with the connected-bus voltage magnitude, inverter current, and apparent-power rating. The apparent-power rating of the grid interface is set to $1.05$ times the maximum power capacity of the connected storage system \citep{nerc2016reactive} \cref{eq:convert_max_rate}. Additionally, the grid interface is assumed to support reactive power independently, with its power capability bounded by the 0.95 leading/lagging power-factor range \citep{federal2016reactive} \cref{eq:convert_reactive_range}. 

\subsubsection{Objective function}
\label{subsubsec:acopf_obj}
The active and reactive power balance constraints in \cref{eq:p_balance,eq:q_balance} are relaxed by introducing load-shedding variables, respectively. These relaxation variables ensure that the optimization remains solvable under different load and renewable scenarios, while their minimum \edit{values quantify} the deviation from AC-feasible operation. Accordingly, given \edit{full-year} scenarios of demand ($\{(P^D_{dt}, Q^D_{dt})\}_{t=1}^{T}, \forall d \in \mathcal{D}$) and renewable availability ($\{\overline{P}^{R}_{rt}\}_{t=1}^{T}, \forall r \in \mathcal{R}$), the objective function is formulated as follows:
{\small
\begin{equation}
\min \quad \sum_{t \in \mathcal{T}}\sum_{d \in \mathcal{D}} S^D_{dt}\alpha^{\mathrm{LS}}_{dt}
\label{eq:ac_obj}
\end{equation}}\noindent where \edit{$S^D_{dt} = \sqrt{(P^D_{dt})^2 + (Q^D_{dt})^2}$ is the apparent load and $\alpha^{\mathrm{LS}}_{dt} = P^{\mathrm{LS}}_{dt}/P^D_{dt}$ is the load-shedding ratio}.
\edit{The decision variables of the full-year AC-OPF comprise the thermal and renewable generator dispatches $(p^{G}_{gt}, q^{G}_{gt})$ and $(p^{R}_{rt}, q^{R}_{rt})$, the bus voltage magnitudes and angles $(v_{nt}, \theta_{nt})$, the line flows $(p_{lt}, q_{lt})$, the load-shedding variables $(P^{\mathrm{LS}}_{dt}, Q^{\mathrm{LS}}_{dt})$, and the storage variables of each type $h$, namely the charging and discharging powers $(p^{\mathrm{CH},h}_{st}, p^{\mathrm{DH},h}_{st})$, the stored energies $E^{h}_{st}$, and the grid-interface variables $(p^{h}_{st}, q^{h}_{st}, q^{\mathrm{int},h}_{st}, I^{h}_{st})$, over all buses, lines, devices, and time periods.}

\subsection{Multi-day security-constrained AC-OPF with LDES}
\label{subsec:SC-AC-OPF}
The role of LDES under $N-1$ contingencies is assessed using a multi-day corrective SC-AC-OPF formulation. \edit{The purpose of this formulation is to quantify the extent to which storage substitutes for the additional capacity that would otherwise be required to guarantee $N-1$ security.} The horizon $\mathcal{T}=\{t_i,\ldots,T\}$ starts at the contingency occurrence time $t_i$ in the full-year horizon and consists of a contingency period $\mathcal{T}_{c}=\{t_i,\ldots,t_c\}$ followed by a restoration period $\mathcal{T}_{r}=\{t_c+1,\ldots,T\}$. This study considers a contingency set $\mathcal{K}=\{1,\ldots,N_c\}$ comprising only line-outage contingencies, together with the base case ($k=0$) without contingency.

Unlike the full-year AC-OPF \cref{eq:ac_obj}, in which load shedding is allowed, the SC-AC-OPF considers only LDES configurations that achieve zero load shedding under the scenario set. Accordingly, the load-shedding \edit{terms} $(P^{\mathrm{LS}}_{dt}, Q^{\mathrm{LS}}_{dt})$ are omitted from the power balance equations in \cref{eq:p_balance,eq:q_balance}. Given the base case and contingency cases $k \in \mathcal{K}$, the resulting equality and inequality constraints of the multi-period AC-OPF with LDES (\cref{subsec:full_year_opf}) are summarized as follows:
{\small 
\begin{subequations} 
\begin{align} 
& g_0(x_{t,0},u_{t,0}) = 0, \quad \forall t \in \mathcal{T} \label{eq:base_eq_const} \\ 
& h_0(x_{t,0},u_{t,0}) \leq 0, \quad \forall t \in \mathcal{T} \label{eq:base_noneq_const} \\ 
& g_k(x_{t,k},u_{t,k}) = 0, \quad \forall t \in \mathcal{T}, \forall k \in \mathcal{K} \label{eq:cont_eq_const} \\ 
& h_k(x_{t,k},u_{t,k}) \leq 0, \quad \forall t \in \mathcal{T}, \forall k \in \mathcal{K} \label{eq:cont_noneq_const}
\end{align} 
\end{subequations}}\noindent where $g_0(\cdot)$ and $h_0(\cdot)$ denote the equality and inequality constraints of the base-case network without contingency, respectively, while $g_k(\cdot)$ and $h_k(\cdot)$ denote the corresponding equality and inequality constraints under contingency $k$. The vector $u_{t,0}$ denotes the dispatch variables of the base-case network, including generator dispatch $(p^{G}_{gt,0}, q^{G}_{gt,0})$ and storage dispatch $(p^{ES}_{st,0}, q^{ES}_{st,0})$, while $x_{t,0}$ denotes the state variables, including branch-flow variables, bus-voltage variables, and the energy states of BESS and LDES. The vectors $u_{t,k}$ and $x_{t,k}$ denote the corresponding dispatch and state variables under contingency $k$, respectively.

Given contingency cases $k \in \mathcal{K}$, the corrective redispatch of thermal and renewable generators is constrained by \edit{the active-power corrective factor $\gamma_{\mathrm{corr}}^{p}$ relative to the base-case dispatch during the contingency period,} as follows:
{\small 
\begin{subequations} 
\begin{align} 
& \left|p^G_{gt,k}-p^G_{gt,0}\right| \leq \gamma_{\mathrm{corr}}^{p} \cdot P^{\mathrm{ramp}}_g \quad \forall g \in \mathcal{G},\ t \in \mathcal{T_c}, 
\label{eq:gen_p_redis}\\
& \left|p^R_{rt,k}-p^R_{rt,0}\right| \leq \gamma_{\mathrm{corr}}^{p} \cdot P^{\mathrm{ramp}}_r \quad \forall r \in \mathcal{R},\ t \in \mathcal{T_c} 
\label{eq:gen_r_redis}
\end{align} 
\end{subequations}}

\begin{figure}[!tp] 
    \centering 
    \includegraphics[width=0.7\linewidth]{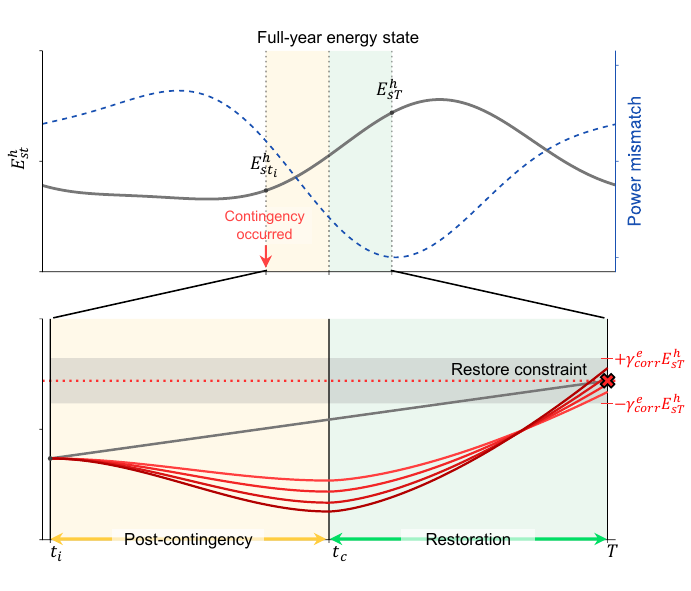}
    \caption{Energy-state coupling between full-year AC-OPF and SC-AC-OPF}
    \label{fig:scopf_formulation} 
\end{figure}

The seasonal energy state of LDES is \edit{required} to be restored during the restoration period following the contingency period. The annual energy-state trajectory $\{E^{\mathrm{h}}_{st}\}_{t=1}^{8760}$ is first obtained by solving the full-year multi-period AC-OPF in \cref{subsec:full_year_opf}. Given the target horizon $\mathcal{T}$, the corresponding energy states at the initial and terminal times $(E_{st_i}^h,E_{sT}^h)$ are defined as the nominal initial and terminal energy states, respectively, thereby coupling the full-year AC-OPF trajectory with the base and contingency cases of the SC-AC-OPF, as shown in \cref{fig:scopf_formulation}. Specifically, the initial and terminal energy states of the base case are enforced to match the corresponding nominal state, as described in \cref{eq:e_state_init_base_match,eq:e_state_end_base_match}. For each contingency case $k$, the initial energy state is also fixed to the nominal initial energy state \cref{eq:e_state_init_cont_match}. \edit{During} the contingency period $\mathcal{T}_c$, additional power is discharged from the storage to support corrective redispatch \cref{eq:gen_p_redis,eq:gen_r_redis}. Accordingly, unlike the base case, the terminal energy state under a contingency may not be fully restored to the nominal state within the restoration period $\mathcal{T}_r$. Thus, as described in \cref{eq:e_state_end_cont_match}, the deviation of the terminal state from the nominal state is constrained by the energy-state corrective factor $\gamma_{\mathrm{corr}}^{e}$.

{\small 
\begin{subequations} 
\begin{align} 
& E^{\mathrm{h}}_{st_i,0} = E^{\mathrm{h}}_{st_i} 
\quad \forall s \in \mathcal{S}, \ h \in \mathcal{H},
\label{eq:e_state_init_base_match}\\
& E^{\mathrm{h}}_{sT,0} = E^{\mathrm{h}}_{sT} 
\quad \forall s \in \mathcal{S}, \ h \in \mathcal{H},
\label{eq:e_state_end_base_match}\\
&E^{\mathrm{h}}_{st_i,k} = E^{\mathrm{h}}_{st_i} 
\quad \forall s \in \mathcal{S}, \ h \in \mathcal{H}, \ k \in \mathcal{K},
\label{eq:e_state_init_cont_match}\\
&\left|E^h_{sT,k}-E^{\mathrm{h}}_{sT}\right| \leq \gamma_{\mathrm{corr}}^{e} \cdot E^{\mathrm{h}}_{sT} \quad \forall s \in \mathcal{S}, \ h \in \mathcal{H}, \ k \in \mathcal{K} 
\label{eq:e_state_end_cont_match}
\end{align} 
\end{subequations}}

The objective function is formulated to minimize the total generation of the base case as follows:
\begin{equation}
\min \quad \sum_{t \in \mathcal{T}} \sum_{g \in \mathcal{G}} p^{G}_{gt,0}
\label{eq:sc_opf_obj}
\end{equation}

\subsection{Uncertainty modeling of load and renewable generation}
\label{subsec:Uncertainty-modeling}
A Gaussian copula method \citep{nelsen2006introduction} is employed to model the uncertainty of renewable generation and load. In this study, the wind speed is \edit{used} as the renewable resource input, and \edit{the} renewable power availability $\overline{P}^{R}_{rt}$ in \cref{eq:renewable_p_capacity_bound} is calculated from the wind speed via wind-turbine power function \citep{wan2010development,kim2023revealing}. Instead of directly modeling the full-year 8,760-hour profile, historical annual wind speed is first decomposed into trend, \edit{periodic (in time-series terminology, seasonal)}, and residual components using locally estimated scatterplot smoothing \citep{seabold2010statsmodels}.

Given the spatial coordinates of renewable generator $r \in \mathcal{R}$ and historical year $yr$, the wind speed $w^{yr}_{rt}$ is decomposed as follows:
\begin{equation}
    w^{yr}_{rt}
    =
    \tau^{yr}_{rt}
    +
    s^{yr}_{rt}
    +
    \epsilon^{yr}_{rt},
    \qquad t \in \{1,\ldots,8760\},
    \label{eq:wind_decomposition}
\end{equation}
\noindent
where $\tau^{yr}_{rt}$ is the trend component, $s^{yr}_{rt}$ is the \edit{periodic component, capturing daily and weekly cycles}, and $\epsilon^{yr}_{rt}$ is the residual component.

\begin{figure}[!tp] 
    \centering 
    \includegraphics[width=0.8\linewidth]{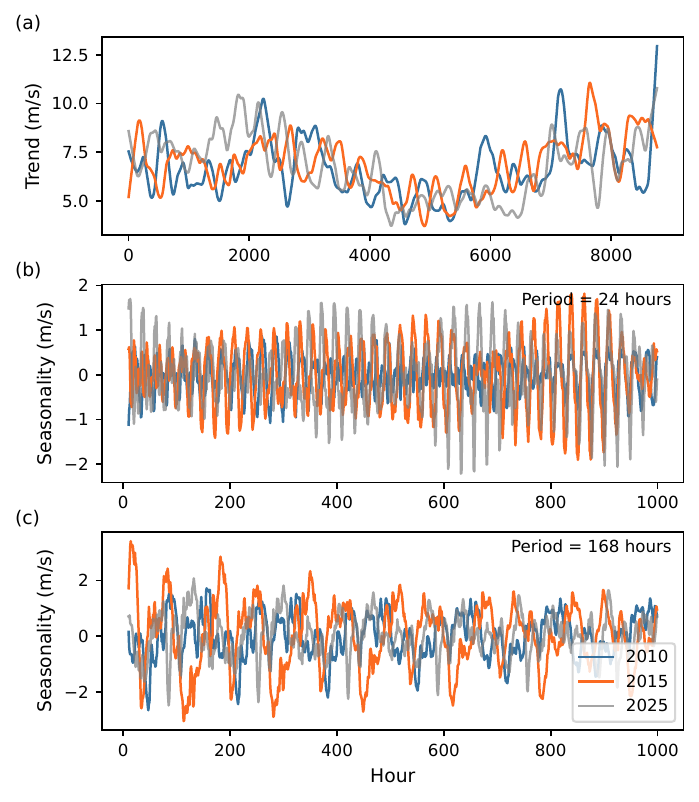}
    \caption{Time-series decomposition of wind speed data across different years: (a) trend component, 
    (b) \edit{periodic} component with a 24-hour period, and (c) \edit{periodic} component with a 168-hour period.}
    \label{fig:wind_decompose} 
\end{figure}

The trend and \edit{periodic} components are preserved to retain the long-term temporal structure of the historical data, while the annual residual component is reshaped into daily 24-hour vectors and used to fit the Gaussian copula model. This decomposition-based approach allows the copula model to capture only residual uncertainty while preserving the long-term trend and cyclic behavior of renewable data, which can vary significantly across historical years, as shown in \cref{fig:wind_decompose}. 

Full-year wind speed scenarios $\tilde{w}^{yr}_{rt}$ are then generated by combining the sampled residuals from the fitted copula model with the extracted trend and \edit{periodic} components, and converted into renewable power availability $\overline{P}^{R,m}_{rt}$ used in \cref{eq:renewable_p_capacity_bound}. The same procedure, from time-series decomposition to copula-based sampling, is applied to the load profile $P^D_{dt}$ and $Q^D_{dt}$ at each load node $d$, used in \cref{eq:p_balance,eq:q_balance}. \edit{A statistical evaluation} of the synthetic scenarios of both wind speed and load is provided in {\hypersetup{linkcolor=red}\cref{subsec:uncertainty_validation}}.

{\small
\begin{subequations}
\begin{align}
    & \tilde{r}_{rt}
    \sim
    \mathcal{C}^{w}_{r},
    \qquad r \in \mathcal{R}, \; t \in \{1,\ldots,8760\}, 
    \label{eq:gc_sample} \\
    & \tilde{w}^{yr}_{rt} 
    =\tau^{yr}_{rt}
    +s^{yr}_{rt}
    +\tilde{r}_{rt},
    \qquad r \in \mathcal{R}, \; t \in \{1,\ldots,8760\}, 
    \label{eq:recover_wind} \\
    & \overline{P}^{R}_{rt}
    =f\left(\tilde{w}^{yr}_{rt}\right),
    \qquad r \in \mathcal{R}, \; t \in \{1,\ldots,8760\}.
    \label{eq:wind_power_conversion}
\end{align}
\end{subequations}}\noindent where $\mathcal{C}^{w}_{r}$ denotes the fitted Gaussian copula model for wind speed residuals of renewable generator $r$, \edit{$\tilde{r}_{rt}$ denotes the residual sampled from this model,} $f(\cdot)$ denotes \edit{the} wind-turbine power function, \edit{whose details are described in \citet{wan2010development,kim2023revealing}}.

\subsection{Case study setup}
\label{subsec:implementation}
\begin{figure}[!tp] 
    \centering 
    \includegraphics[width=1.0\linewidth]{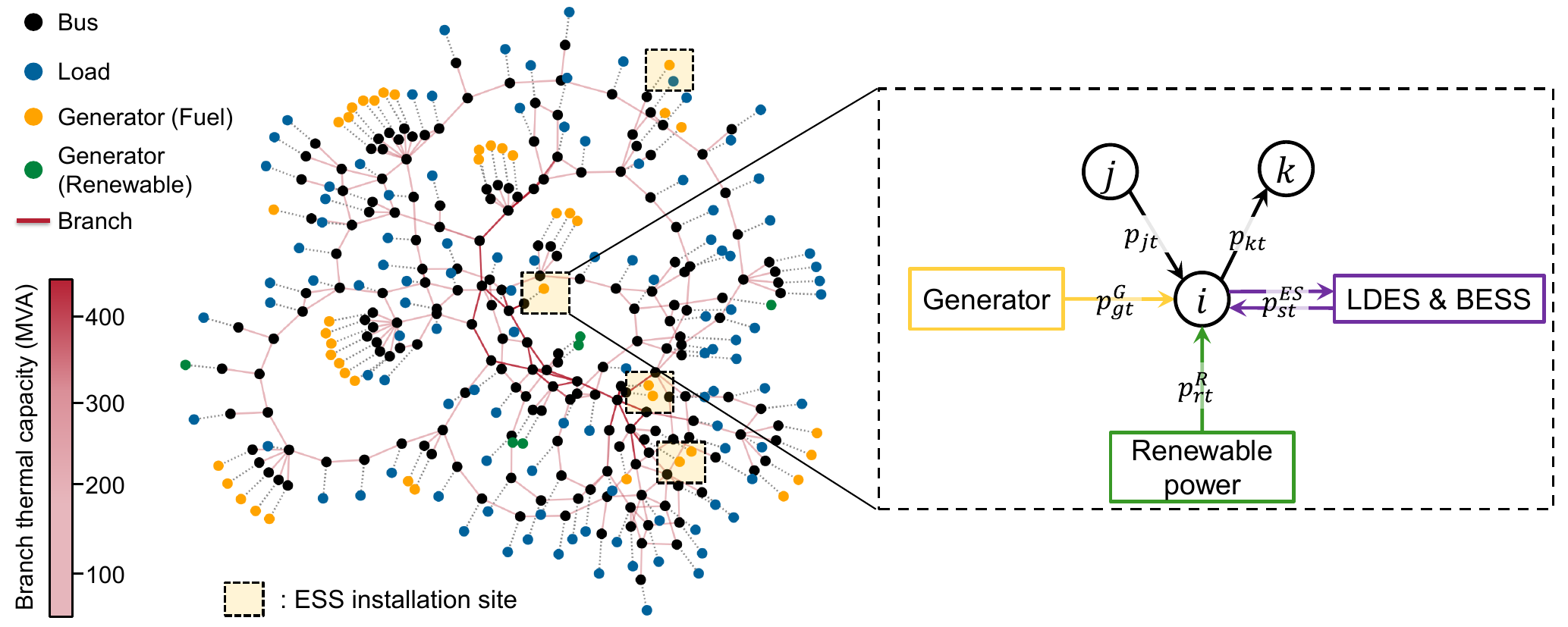}
    \caption{Modified 200-bus network with renewable generation and ESS}
    \label{fig:network_diagram} 
\end{figure}

This study uses a synthetic Illinois network with bus-level load profiles, geographical information, and generator fuel-type data \citep{birchfield2016grid,li2018load}. The network consists of 200 buses and 49 generators, including five renewable-based and 44 thermal generators. Among the 44 thermal generator buses, the top five sites are selected based on the maximum allowable power flow of their adjacent transmission lines and used as representative test sites, as shown in \cref{fig:network_diagram}. To simulate high-renewable-penetration conditions, additional renewable generation is assumed at the selected sites by adjusting the thermal generator capacity $P^{\max}_g$ in \cref{eq:gen_p_bound}. \edit{This modification emulates a future decarbonized grid, in which part of the existing thermal fleet is retired and replaced by renewable generation of matching capacity at the same sites. The substitution ratio of 60\% is consistent with the renewable share of electricity generation projected for 2030 under the IEA Net Zero Emissions scenario \citep{iea2021netzero}.} Given \edit{a} capacity ratio \edit{of} $\alpha=0.6$, the thermal generator capacity is reduced to $(1-\alpha)P^{\max}_g$, while the additional renewable power is added with a capacity of $\alpha P^{\max}_g$. BESS and LDES with different power capacities are then co-located at the selected sites. \edit{The resulting base network, in which thermal capacity is retired without any compensating investment, is not intended as a realistic operating state; it is the counterfactual that quantifies the reliability gap created by thermal retirement under high renewable substitution, which the storage configurations must close. All reported results are comparisons between configurations under identical scenarios.}

For the full-year AC-OPF optimization, 100 scenarios are generated for renewable generation and load, based on \cref{subsec:Uncertainty-modeling}. Since load profiles are available only for 2017 \citep{birchfield2016grid,li2018load}, the load scenarios are generated from the 2017 data, whereas renewable scenarios are generated using historical renewable resource profiles retrieved from the NASA POWER API \citep{sparks2018nasapower} for 2010--2025 based on the geographical coordinates of the target renewable generator. Full-year scenarios are reconstructed by uniformly sampling historical years and combining sampled residual vectors with the trend and \edit{periodic} components of each sampled year \cref{eq:wind_decomposition}.

For the SC-AC-OPF optimization, the nominal energy-state trajectory is obtained by solving the full-year AC-OPF using historical load and renewable profiles for 2020. To cover a wide range of contingency occurrence times, including peak-load periods, the SC-AC-OPF is iteratively solved with different contingency occurrence times throughout the full-year horizon. Specifically, each SC-AC-OPF considers a 24-hour contingency period $\mathcal{T}_c$ followed by a 24-hour restoration period $\mathcal{T}_r$. The contingency occurrence time $t_i$ is shifted at 7-day intervals throughout the target year of 2020. Additionally, this study considers only line-outage contingencies to focus on the impact of LDES on secure network operation associated with power flow. Due to the computational complexity of the problem, line-outage contingencies are limited to the top 50\% of transmission lines ranked by its thermal capacity (\cref{fig:network_diagram}).

For optimization, this study leverages the GPU-compatible nonlinear power flow model ExaModelsPower.jl \citep{johnson2025examodelspower} with interior-point solver MadNLP.jl \citep{shin2024accelerating}. The termination tolerance and maximum iterations of the solver are set to $10^{-6}$ and 500, respectively. Computations are performed on a computing server equipped with an Intel Xeon Platinum 8480C CPU with 56 cores and 112 threads, up to 3.8 GHz, 2.01 TB of RAM, and a single NVIDIA H200 GPU with 140 GB of memory and 700 W TDP.

\section{Results and discussion}
\label{sec:results}
\subsection{Statistical evaluation of synthetic scenarios}
\label{subsec:uncertainty_validation}
As described in \cref{subsec:Uncertainty-modeling}, the Gaussian copula method is employed to construct full-year scenario sets of wind speed and load for full-year AC-OPF optimization. Before conducting the optimization, 100 sampled scenarios are \edit{compared} against the corresponding historical datasets using percentile values of the hourly profiles and autocorrelation coefficients at different time lags.

\begin{figure}[!tp] 
    \centering 
    \includegraphics[width=1.0\linewidth]{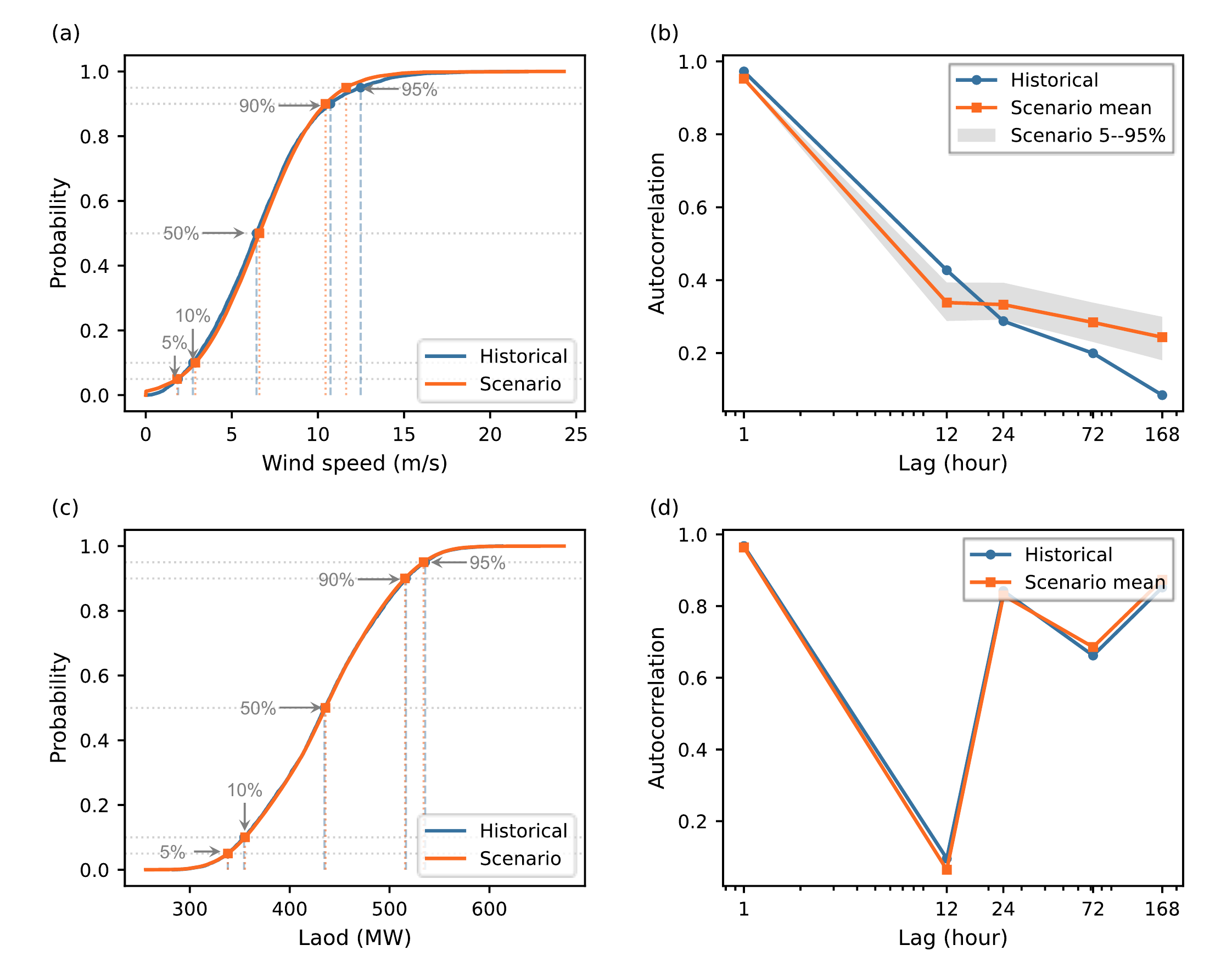}
    \caption{Comparison of statistical feature between sampled scenarios and historical data: (a) cumulative distribution, (b) autocorrelation factors of wind-speed, (c) cumulative distribution, (d) autocorrelation factors of load demand.}
    \label{fig:scenario_valid} 
\end{figure}

\Cref{tab:scenario_validation} summarizes the mean error and standard deviation of these metrics across historical years and spatial locations, showing that the sampled full-year scenarios closely reproduce the marginal distributions and temporal dependence. Specifically, the percentile errors remain within 5\% for wind speed and within 1\% for load demand, while the low autocorrelation errors from 1-hour to 168-hour lags indicate that temporal correlation is well preserved. \Cref{fig:scenario_valid} further illustrates the agreement in hourly-value distributions and autocorrelation trends between the generated and historical profiles.

\begin{table}[!tp]
\centering
\caption{Comparison of statistical metrics between sampled scenarios and historical data}
\label{tab:scenario_validation}
\resizebox{\linewidth}{!}{
\begin{tabular}{lccccc}
\toprule
\multicolumn{6}{l}{\textbf{Error in selected percentile values (\%)}} \\
\midrule
Data & 5th & 10th & 50th & 90th & 95th \\
\midrule
Wind speed 
& $4.5 \pm 2.5$ 
& $5.3 \pm 2.1$ 
& $1.6 \pm 0.8$ 
& $1.7 \pm 1.2$ 
& $4.7 \pm 1.8$ \\
Load demand
& $0.2 \pm 0.3$ 
& $0.1 \pm 0.1$ 
& $0.3 \pm 0.5$ 
& $0.4 \pm 0.6$ 
& $0.5 \pm 0.7$ \\
\midrule
\multicolumn{6}{l}{\textbf{Absolute error of autocorrelation factor}} \\
\midrule
Data & Lag 1 & Lag 12 & Lag 24 & Lag 72 & Lag 168 \\
\midrule
Wind speed 
& $0.02 \pm 0.001$  
& $0.10 \pm 0.02$ 
& $0.03 \pm 0.03$ 
& $0.09 \pm 0.02$ 
& $0.11 \pm 0.03$ \\
Load demand 
& $0.002 \pm 0.002$  
& $0.03 \pm 0.02$ 
& $0.01 \pm 0.01$ 
& $0.02 \pm 0.02$ 
& $0.02 \pm 0.02$ \\
\bottomrule
\end{tabular}
}
\vspace{0.5ex}
\begin{minipage}{0.98\columnwidth}
\scriptsize
\textit{Note:} Values are reported as mean $\pm$ standard deviation across historical years and across renewable-generator sites for wind speed or load-demand nodes for demand.
\end{minipage}
\end{table}

\subsection{Full-year network feasibility improvement with LDES}
\label{subsec:full_year_result}
\begin{figure}[!tp]
\centering
\includegraphics[width=1.0\linewidth]{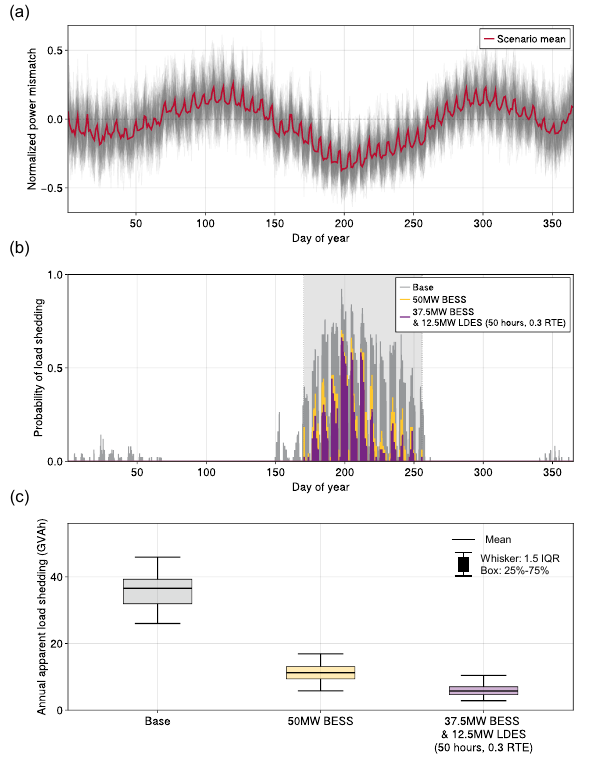}
\caption{Network load shedding under renewable power and load uncertainty: (a) annual imbalance between normalized renewable availability and load across scenarios, (b) probability of hourly load-shedding events across scenarios, and (c) distribution of annual apparent load shedding for the base network without ESS, with standalone BESS, and with hybrid BESS--LDES installation.}
\label{fig:hourly_shedding}
\end{figure}

Full-year load-shedding minimization was conducted for the 200-bus target network under 100 synthetic renewable and demand scenarios. \edit{In this setting, renewable generators matching the retired thermal capacity are already installed (\cref{subsec:implementation}), while no storage capacity is installed yet; the storage configurations are introduced in the subsequent analyses.} \Cref{fig:hourly_shedding} (a) shows the hourly imbalance between renewable availability and demand, highlighting negative power imbalances during the early-year period (days 0--120), summer period (days 150--275), and winter period (days 340--360). In the base network without ESS, load-shedding events occur over a wide range of the year and are closely aligned with these negative power-imbalance periods, as shown in \cref{fig:hourly_shedding} (b). Moreover, \edit{the annual apparent load shedding of the base network, without any ESS, ranges from 25 to 45 GVAh (gigavolt-ampere-hours) across the 100 scenarios (\cref{fig:hourly_shedding} (c)); because the optimization minimizes shedding for each scenario, with generator dispatch as the only source of flexibility, these values are the minimum attainable shedding, not an accumulated energy imbalance}. These results indicate that network operation under high renewable penetration induces \edit{prolonged periods of negative power imbalance}, thereby leading to load-shedding events across the full-year period. \edit{Because even the minimum attainable shedding is strictly positive, this load shedding is inevitable for the storage-less network under the considered scenarios: no operational strategy can avoid it, and only additional capacity can. This shedding quantifies the reliability gap that thermal retirement creates in the counterfactual base network (\cref{subsec:implementation}); the following analyses examine to what extent storage, in place of additional generation or transmission capacity, closes this gap.}

% [Q for first author, SS 2026-09-06] report per configuration: (a) fraction of hours with nonzero shedding, (b) share of annual shedding in the worst 1% of hours -- I expect shedding to be confined to tail events for the storage-equipped cases; if the data does not support this, the case study may need revision
In contrast, installation of an ESS with a total power capacity of 50 MW eliminates load-shedding events during the early-year and winter periods (\cref{fig:hourly_shedding} (b)). Additionally, an ESS-equipped network substantially reduces the probability of hourly load-shedding events during \edit{the} summer peak period compared with the base network. Furthermore, for the same total ESS power capacity of 50 MW, the hybrid BESS--LDES configuration, consisting of a 37.5 MW BESS (4~h and 0.77 round-trip efficiency (RTE)) and a 12.5 MW LDES (50~h duration and 0.3 RTE), further reduces the probability of load-shedding events compared with the standalone BESS configuration. In terms of annual load shedding, the hybrid BESS--LDES configuration achieves an average reduction of 83.0\% relative to the base network, compared with a 68.5\% reduction achieved by the standalone BESS configuration (\cref{fig:hourly_shedding} (c)).

\begin{figure}[!tp]
\centering
% [SS 2026-09-06, for JK] trim the x-axis of this figure appropriately
\includegraphics[width=1.0\linewidth]{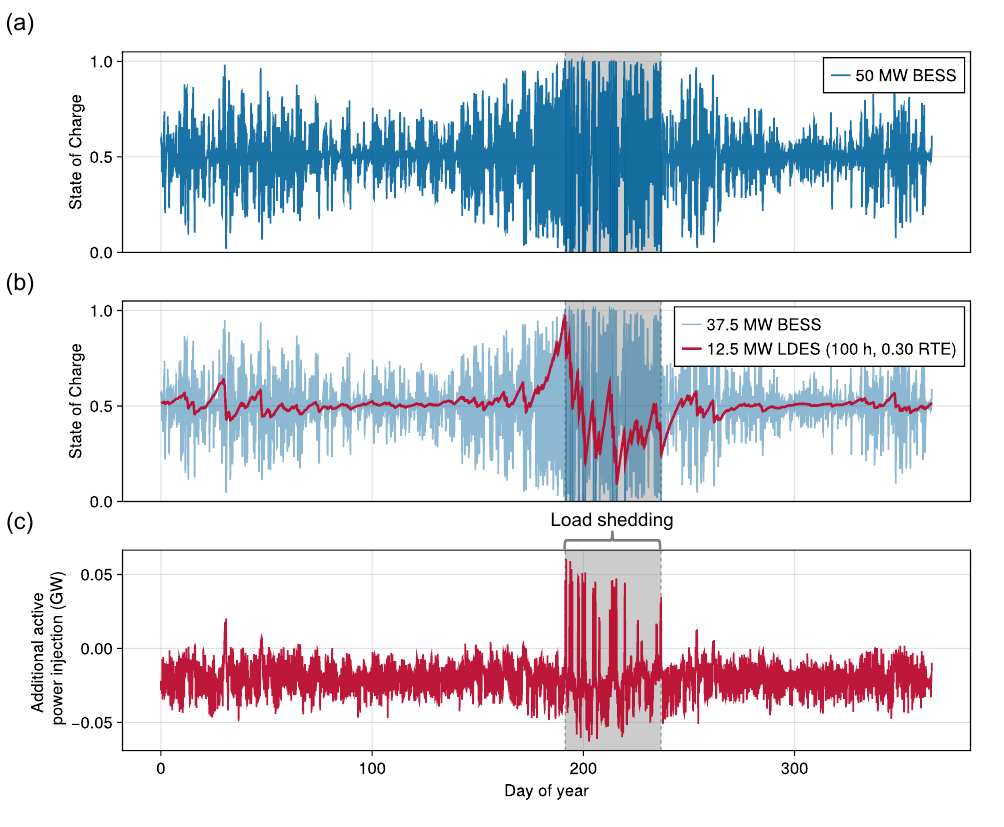}
\caption{State of Charge (SoC) profiles of BESS and LDES: (a) SoC of 50 MW standalone BESS, (b) SoC of 12.5 MW LDES co-installed with 37.5 MW BESS, and (c) active power dispatch difference between hybrid BESS--LDES and standalone BESS.}
\label{fig:SoC_profile}
\end{figure}

The annual state of charge (SoC) profiles of the two ESS configurations further explain the benefit of LDES in reducing load-shedding, as shown in \cref{fig:SoC_profile}. Owing to the short duration of the BESS (4~h), the BESS in both configurations shows similar cyclic behavior. Benefiting from its high charge and discharge efficiency (0.95, 0.90), the BESS performs more than 900 annual cycles on average across the scenarios despite its limited duration (\cref{fig:SoC_profile} (a) and (b)). % [Q for first author] 3 cycles/day is the physical maximum for a 4-h device (24h/8h) -- verify from the data whether this is the daily MEAN over days 150-275 or the saturated maximum on stressed days, and consider stating the no-degradation-cost assumption (the optimizer cycles for free)
Specifically, during the summer period (days 150--275), the BESS performs 3 cycles per day with full charge and discharge on average, thereby reducing \edit{load shedding} caused by peak load and renewable intermittency. In contrast, the LDES does not rely on frequent cycling due to its low RTE of 0.3. Instead, its longer duration of 50~h enables energy shifting for seasonal balancing. As shown in \cref{fig:SoC_profile} (b), the SoC of the LDES gradually reaches unity before the summer period and decreases to its minimum SoC of 0.1 during this period. Accordingly, the LDES provides its maximum active power dispatch (0.05 GW) to the network primarily during the summer load peak period, while it is charged during the remaining periods of the year (\cref{fig:SoC_profile} (c)).

Beyond the fixed LDES parameter setting, a sensitivity analysis was conducted to identify which LDES parameters most strongly affect network feasibility. The analysis varies LDES power capacity, duration, and RTE. To isolate the contribution of LDES to the improvement in network feasibility, 25\% of a given total ESS power capacity is allocated to LDES, while the remaining 75\% is allocated to co-installed BESS. For each LDES parameter setting, the full-year AC-OPF is solved across the scenario set.

\begin{figure}[!tp]
\centering
\includegraphics[width=1.0\linewidth]{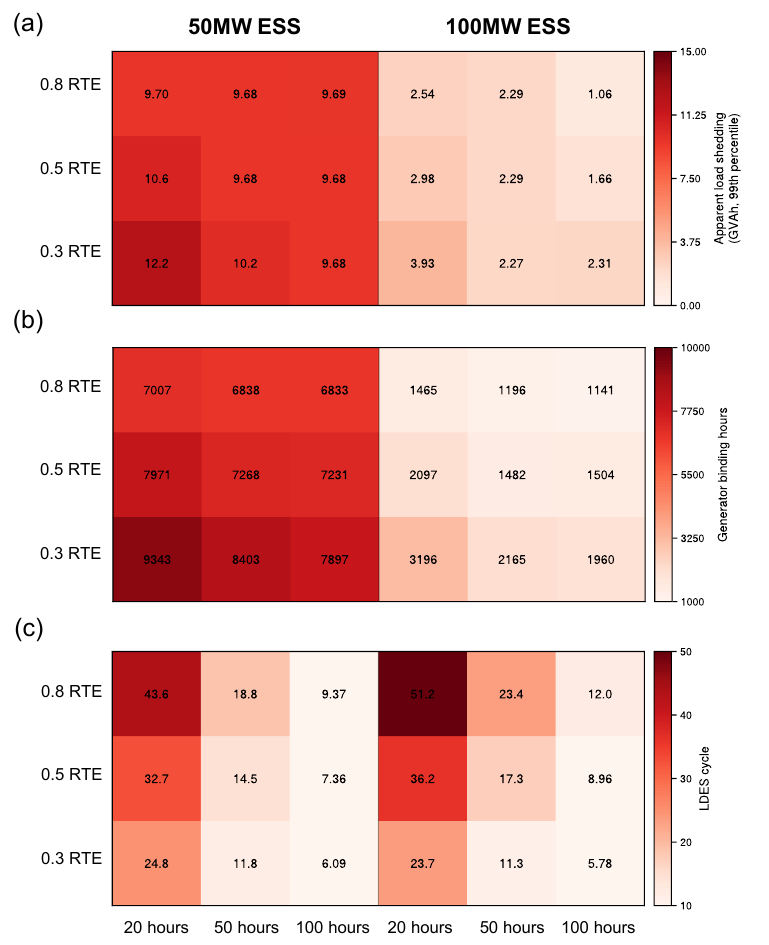}
\caption{Network feasibility under different LDES parameters and capacities: (a) 99th-percentile annual apparent load shedding across scenarios, (b) average total generator binding hours across scenarios, and (c) average annual cyclic number of LDES across scenarios}
\label{fig:grid_search}
\end{figure}

For the 20~h duration setting, increasing RTE reduces load shedding for both the 50 MW and 100 MW total ESS capacity cases, as shown in \cref{fig:grid_search} (a). This trend is consistent with the hourly storage energy balance \cref{eq:storage_energy_update}, in which a higher RTE reduces energy losses during charge--discharge cycles. In particular, under the 20 h duration setting, the LDES is required to operate with 25--48 cycles per year, as shown in \cref{fig:grid_search} (c), indicating that cycling efficiency directly affects its ability to buffer renewable intermittency. As a result, increasing RTE from 0.3 to 0.8 reduces the 99th-percentile annual apparent load shedding by 20.4\% and 35.4\% for the 50 MW and 100 MW power capacity cases, respectively. In addition, higher RTE reduces generator binding hours, defined as the hours in which generators operate above 99\% of their maximum power capacity, as shown in \cref{fig:grid_search} (b). These results indicate that, under short-duration settings, \edit{high RTE not only reduces load shedding but also alleviates the binding operation of existing generators} over the full-year horizon.

However, once the LDES duration exceeds 50~h, further increases in RTE or duration do not lead to a noticeable reduction in load shedding. This plateau is observed consistently from 50 h to 100 h \edit{durations} (\cref{fig:grid_search} (a)). Therefore, the limiting factor shifts from energy capacity to \edit{instantaneous power availability}, particularly during peak-load periods (\cref{fig:hourly_shedding}). As a result, increasing the total ESS power capacity from 50 MW to 100 MW produces a much larger reduction in load shedding than further increasing duration or RTE. For example, under the 50~MW and 50~h duration setting, increasing RTE from 0.3 to 0.8 reduces load shedding by only 5.1\%, whereas increasing the power capacity to 100 MW achieves a 76.3\% reduction across all RTE settings.

To further explain this plateau behavior over a 50 h duration, constraint binding hours are analyzed during load-shedding event periods. In this study, binding hours are defined as time periods in which the relevant constraint reaches its lower or upper bound within a 1\% tolerance. The binding constraints are classified into \edit{those associated with the bus connected to the LDES}, including line-flow and generator active/reactive power limits in \cref{eq:line_limit,eq:gen_p_bound,eq:gen_q_bound,eq:renewable_p_capacity_bound,eq:renewable_q_bound}; LDES power limits in \cref{eq:storage_charge_bound,eq:storage_discharge_bound}; inverter apparent-power limits in \cref{eq:converter_flow_limit}; and LDES storage capacity limits in \cref{eq:storage_energy_bound}.

\begin{figure}[!tp]
\centering
\includegraphics[width=1.0\linewidth]{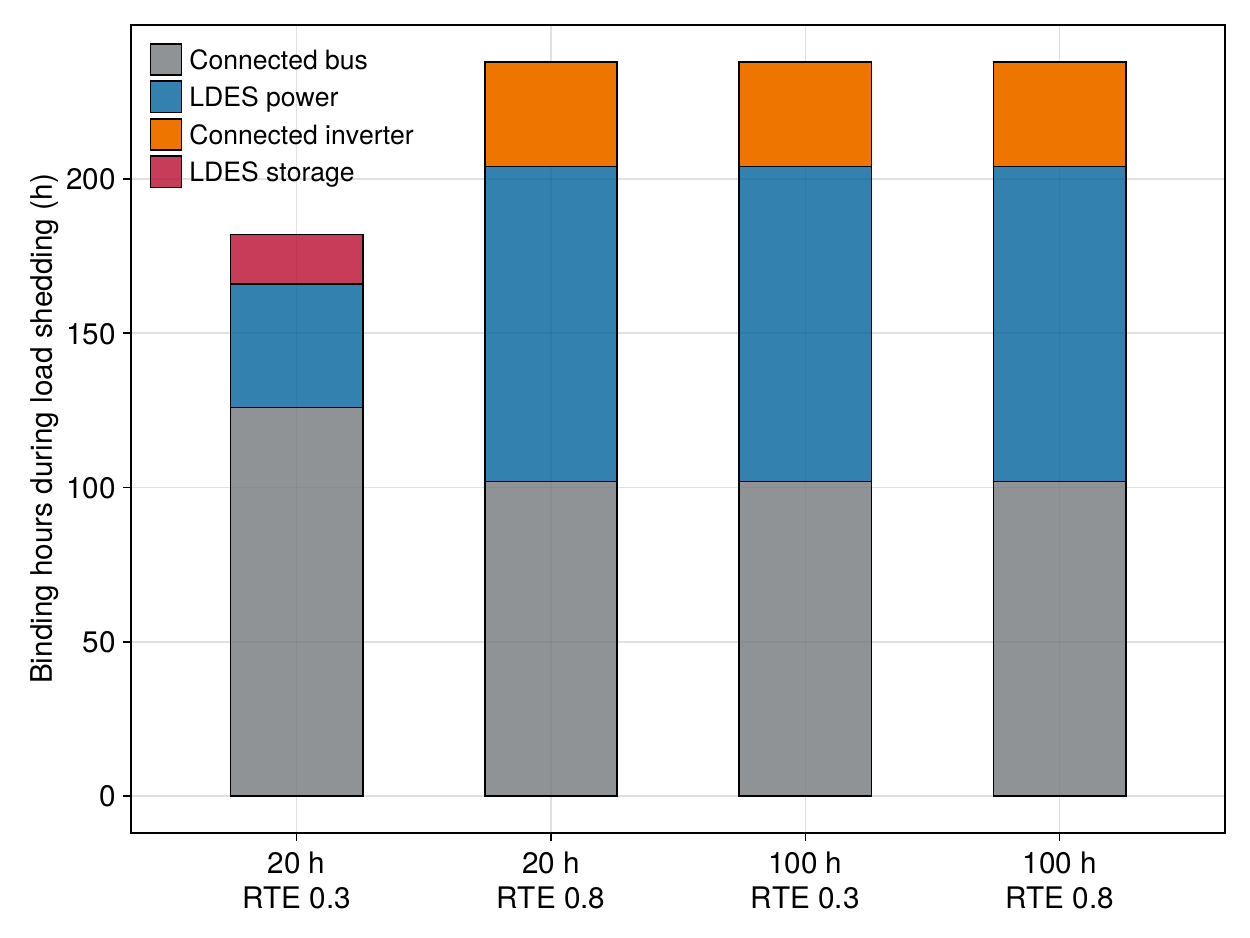}
\caption{Constraint binding hours during load-shedding event periods under different LDES parameter settings.}
\label{fig:binding_hours}
\end{figure}

\Cref{fig:binding_hours} reports the averaged binding hours for four representative LDES settings during load-shedding events. For the 20 h duration case with the lowest RTE of 0.30, storage capacity binding hours are observed, indicating that $E_{st}^{\mathrm{LD}}$ frequently reaches either its lower or upper energy bound during load-shedding events. In this case, the limited storage capacity and low RTE jointly degrade the performance of LDES. By contrast, when either RTE is increased or duration is extended, storage capacity binding hours disappear, while LDES power limits and inverter limits become the dominant binding hours. This indicates that once sufficient energy capacity is secured, the bottleneck shifts from energy availability to instantaneous power supply capability. Although all four representative settings still exhibit binding hours associated with the connected bus, the comparison suggests that the plateau in load-shedding reduction \edit{beyond 50 h durations} observed in \cref{fig:grid_search} (a) is primarily caused by power-rating and converter limits rather than by insufficient storage capacity or RTE. \edit{These binding patterns pinpoint the infeasibility scenario that extending the storage duration resolves: during prolonged scarcity events, short-duration storage exhausts its stored energy before the event ends, whereas extended durations keep stored energy available across the entire event.} 

\begin{figure}[!tp]
\centering
\includegraphics[width=1.0\linewidth]{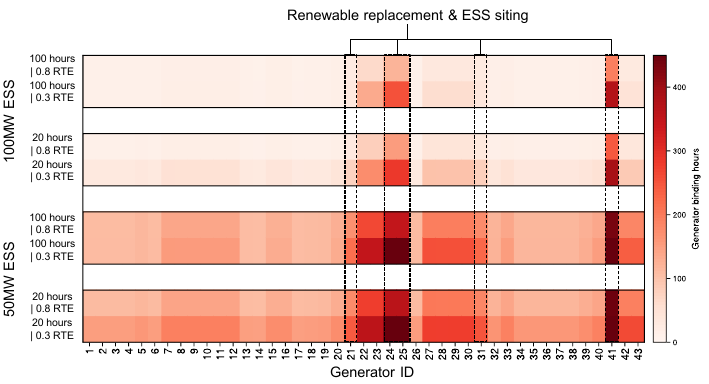}
\caption{Individual generator binding hours under different LDES parameter and power capacity settings}
\label{fig:gen_binding}
\end{figure}

% [Q for first author] "43 existing generators" below vs "44 thermal generators" in implementation details -- which is correct?
Despite the saturation in load-shedding reduction, generator binding hours continue to decrease with increasing RTE, duration, and power capacity (\cref{fig:grid_search} (b)). To examine this effect across the network, the binding hours of individual generators are compared under different LDES parameter settings, as shown in \cref{fig:gen_binding}. Among the 43 existing generators, generators 21, 24, 25, 31, and 41 are co-located with additional renewable generators, as described in \cref{fig:network_diagram} and \cref{subsec:implementation}. Accordingly, compared with the remaining generators, these generators exhibit higher binding hours under high renewable penetration. The results \edit{demonstrate} that installing ESS at the corresponding generator buses, together with increases in both RTE and ESS power capacity, reduces not only the binding hours at those buses but also those of the remaining generators. Specifically, for the 50 MW and 100 h setting, increasing RTE from 0.3 to 0.8 reduces the average binding hours of generators at the ESS installation sites by 27.4\%, while reducing those of the remaining generators by 10.3\%. Moreover, increasing the total ESS power capacity to 100 MW reduces the binding hours of most remaining generators to nearly zero and further reduces those of generators at the ESS installation sites by approximately 66.9\% at an RTE of 0.8. Overall, these results demonstrate that LDES installation can directly alleviate generator binding at the corresponding locations while also providing network-wide operational relief through the transmission network.

\subsection{Curtailment reduction and nominal operation with LDES}
% [Q for first author, SS 2026-09-06] we should add a comparative study against short-duration storage in which the dispatch is cost-based; the current analysis contains no short-duration comparison, and cost-based dispatch is needed to evaluate the curtailment impact realistically.
\label{subsec:generation_analysis}
Beyond the minimum annual load shedding of 1.06 GVAh at an ESS power capacity of 100 MW shown in \cref{fig:grid_search}, a further increase in ESS power capacity can achieve zero load shedding across the scenario set, as demonstrated by a total ESS power capacity of 150 MW, an RTE of 0.5, a duration of 100 h, and a 25\% share of LDES power capacity. Thus, beyond load-shedding minimization, an additional full-year optimization is conducted to examine \edit{the extent to which the storage absorbs otherwise-curtailed renewable energy and thereby reduces thermal generation}. \edit{Accordingly, the objective is changed} to total generation minimization \cref{eq:sc_opf_obj}, with the load-shedding variables $(P^{\mathrm{LS}}_{dt}, Q^{\mathrm{LS}}_{dt})$ omitted from the power balance equations \cref{eq:p_balance,eq:q_balance}. \edit{Because the load is served in full, minimizing thermal generation absorbs the maximum renewable energy; the objective is therefore equivalent, up to network losses, to minimizing renewable curtailment.} \edit{For comparison, the original network, in which the full thermal fleet is retained and neither renewable substitution nor ESS is applied, is optimized over the same scenario set. In this comparison, load shedding is not permitted in either configuration, so both serve the load in full, and generation refers to the total thermal generation.}

\begin{figure}[!tp]
\centering
\includegraphics[width=0.8\linewidth]{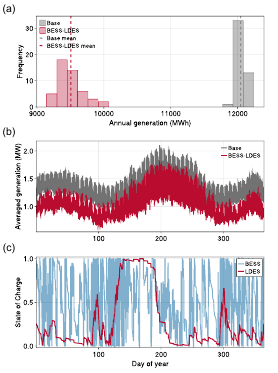}
\caption{Full-year generation comparison between the base network and the network with 150 MW ESS installation: (a) annual generation distribution across the scenario set, (b) hourly generation profiles, and (c) state of charge profiles of the installed BESS and LDES.}
\label{fig:economic_dispatch}
\end{figure}

\Cref{fig:economic_dispatch} (a) shows that the LDES-equipped network with BESS achieves lower annual \edit{thermal} generation than the \edit{original} network across all scenarios, with an average reduction of 21.2\%. \edit{This reduction reflects the combined effect of the renewable substitution and the storage that renders it feasible: the renewables displace the thermal energy, while the storage enables the substituted network to serve the full load.} Specifically, the hourly generation profiles in \cref{fig:economic_dispatch} (b) show that the LDES-installed network reduces generation throughout the full-year horizon rather than solely during specific periods such as the summer peak period (days 150--275), where negative power imbalances are observed (\cref{fig:hourly_shedding}(a)). Similar to the load-shedding minimization results, the generation minimization solution also reveals distinct storage dynamics between BESS and LDES. \Cref{fig:economic_dispatch} (c) shows that LDES supports seasonal energy shifting, particularly by discharging during the summer peak period. Specifically, its SoC reaches unity before the peak period and decreases to a minimum level of 0.1 over the peak period. In contrast, BESS exhibits a high \edit{number of cycles}, supporting short-term energy balancing. 

Overall, this additional analysis on the generation minimization solution demonstrates two key benefits. First, for certain ESS power capacities, LDES installation enables full-year AC feasibility under high renewable penetration across the renewable and load scenarios considered. Second, beyond attaining AC feasibility, the LDES installation with BESS effectively reduces annual generation from existing thermal generators. These two improvements highlight the role of the LDES deployment with BESS in reducing reliance on thermal generation under high renewable penetration.

\subsection{Corrective redispatch and LDES operation under $N-1$ contingencies}
\label{subsec:contingecy_analysis}
% [Q for first author, SS 2026-09-06] for a future revision, the right case-3 analysis is to quantify the capacity substitution directly: (a) without LDES, how much additional capacity (generation, transmission, or short-duration storage) must be built for the contingency problem to become feasible; (b) with LDES, how much that required buildout shrinks. The current feasible/infeasible contrast is the qualitative version of this; the avoided-capacity number should be the headline.
Based on \cref{subsec:full_year_result,subsec:generation_analysis}, the ESS configuration that guarantees full-year network feasibility (150 MW total power capacity, 0.5 RTE, and 100 h LDES duration) is further evaluated under $N-1$ line contingencies using the SC-AC-OPF formulation in \cref{subsec:SC-AC-OPF}. As described in \cref{subsec:implementation}, given 24 h contingency and restoration periods, the problem is independently solved for a series of contingency occurrence times $t_i$, shifted at 7-day intervals throughout the year. The same optimization is performed for the original network without BESS--LDES and standalone BESS installation. No feasible solution is obtained for any of these independently solved problems, whereas the network with BESS--LDES yields feasible solutions for all of them.
\edit{This demonstrates that, under high renewable penetration, storage with sufficient duration restores $N-1$-secure operation without any additional generation or transmission expansion: at equal total power capacity, the standalone-BESS addition yields no feasible solution for any contingency window, whereas the BESS--LDES configuration is feasible for all of them. In this sense, LDES substitutes for the additional generation or transmission capacity that would otherwise be required for secure operation. Moreover, unlike conventional reserve capacity that remains idle outside contingencies, the same LDES capacity is actively used in normal operation, providing \edit{the seasonal balancing and curtailment reduction quantified in \cref{subsec:generation_analysis}}.}

\begin{figure}[!tp]
\centering
\includegraphics[width=0.9\linewidth]{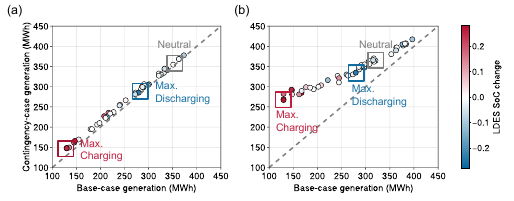}
\caption{Comparison of total generation between the base case and the contingency case during (a) the contingency period and (b) the restoration period. Colors indicate the change in LDES state of charge.}
\label{fig:sc_opf}
\end{figure}

\begin{figure}[!tp]
\centering
\includegraphics[width=0.9\linewidth]{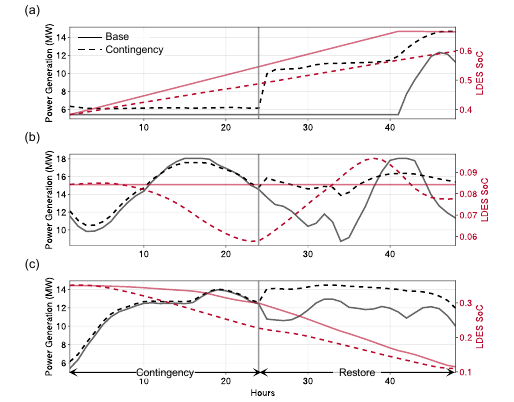}
\caption{Generator redispatch patterns for three representative cases characterized by LDES state-of-charge changes after the restoration period: (a) charging, (b) neutral, and (c) discharging.}
\label{fig:sc_pattern}
\end{figure}

\Cref{fig:sc_opf} compares the average generation across $N-1$ contingency cases with the corresponding base-case (no contingency) generation for all considered contingency occurrence times. During the contingency period, line outages restrict power transfer capability \cref{eq:line_limit}, requiring corrective redispatch of the generators. Even under the 10\% redispatch constraint relative to the hourly base-case generation \cref{eq:gen_p_redis,eq:gen_r_redis}, the BESS--LDES network maintains feasible operation, with contingency-case generation only 5.5\% higher on average than the base-case generation, as shown in \cref{fig:sc_opf} (a). Additional power supplied from LDES reduces the required corrective generator redispatch. \Cref{fig:sc_pattern} shows lower LDES SoC in the contingency case (red dotted line) than in the base case (red solid line) across the three representative operating patterns, further demonstrating the role of LDES as a backup power supply during the contingency period.

In the subsequent restoration period, the \edit{outaged line is restored}, and the LDES energy depleted during the contingency period is recovered toward the terminal energy state of nominal operation (\cref{fig:economic_dispatch} (c)), as enforced by \cref{eq:e_state_end_cont_match}. Consequently, the difference in total generation between the contingency and base cases increases by an average of 60.29 MWh compared with the contingency period. However, the amount of energy recovery depends on the target terminal LDES energy state, resulting in varying generation differences from the base case, as shown in \cref{fig:sc_opf} (b). 

In the charging case (red point in \cref{fig:sc_opf} (b)), the network must recover both the depleted LDES energy and the additional energy required for LDES charging. Accordingly, the maximum-charging case requires 94.8\% higher generation than the base case. \Cref{fig:sc_pattern} (a) further illustrates the hourly generation and LDES operating patterns during the restoration period. In the \edit{nominal} full-year LDES operation (\cref{fig:economic_dispatch} (c)), this charging case corresponds to a contingency occurring before the summer peak period (days 150--275), when the LDES is being charged.

In the neutral case, where the LDES maintains its initial energy state, only the depleted energy needs to be restored (\cref{fig:sc_pattern} (b)). Thus, less generation is required than in the charging case, with a generation gap of only 13.1\%. Similarly, in the LDES discharging case (blue point in \cref{fig:sc_opf} (b)), restoration can be achieved by restricting the LDES discharge rate rather than by substantially increasing generation to further charge the LDES (\cref{fig:sc_pattern} (c)). As a result, in the maximum-discharging case, the generation difference between the contingency and base cases is only 21.4\%. In the \edit{nominal} full-year LDES operation (\cref{fig:economic_dispatch} (c)), this case corresponds to a contingency occurring during the summer peak period, when the LDES is required to continuously discharge to support network operation.

Overall, the \edit{contingency analysis demonstrates} the role of LDES as a backup power supply during contingency periods, satisfying load while limiting corrective generator redispatch to within 10\%. Although additional generation is required, the subsequent 24 h restoration period fully recovers the LDES energy state required for continued nominal operation. Furthermore, obtaining feasible solutions for contingency occurrence times spanning the entire year demonstrates that LDES enables AC-feasible network operation under $N-1$ contingencies across the considered renewable generation and load conditions throughout the year.

\clearpage

\section{Conclusion}
\label{sec:conclusion}
\edit{This study investigated the reliability value of LDES in high-renewable grids against extreme events, encompassing Dunkelflaute-driven scarcity and $N-1$ line contingencies. For a modified 200-bus synthetic network, a full-year multi-period AC-OPF formulation, combined with tail-event characterization through copula-based scenarios and the 99th-percentile load-shedding metric, holistically evaluates network feasibility and reliability, and its seasonally anchored SC-AC-OPF extension evaluates secure operation under contingencies. The results demonstrate that LDES effectively provides the reliability that the renewable-substituted grid otherwise lacks.}

The full-year optimization reveals that, compared with the base network, LDES installation reduces both the occurrence of load-shedding events and total load shedding. Distinct \edit{operating patterns of the two storage types} underlie this improvement, with LDES primarily shifting energy \edit{at the seasonal time scale toward prolonged scarcity periods}, while the co-installed BESS mainly supports daily balancing. Sensitivity analysis further identifies how LDES parameter settings affect network feasibility. Increasing RTE is effective in reducing load shedding, but its benefit saturates beyond approximately 50 h of duration. This indicates that reducing cycling energy losses is critical at limited storage durations, whereas once sufficient energy capacity is available, \edit{the storage power rating becomes the binding constraint, particularly} during critical periods such as the summer load peak. \edit{For configurations with adequate capacity to serve the full load, LDES further reduces renewable curtailment, thereby minimizing thermal generation.}

Under $N-1$ line contingencies, the network without BESS--LDES or with standalone BESS yields no feasible solution, whereas LDES enables corrective operation with restricted generator redispatch and full recovery of the depleted energy within the subsequent 24 h restoration period. \edit{Thus, LDES deployment can substitute for the additional generation or transmission expansion that would otherwise be required to maintain $N-1$-secure operation under high renewable penetration; unlike such reserve capacity, which remains largely idle outside contingencies, the LDES stays actively utilized in normal operation.} Feasibility across all considered contingency occurrence times \edit{demonstrates} that LDES can support network operation under different renewable generation, load, and storage conditions throughout the year. The required corrective redispatch and restoration generation, however, vary with contingency timing and the corresponding LDES energy state, indicating that contingency response depends strongly on seasonal LDES operation. \edit{LDES thus provides value against both prolonged scarcity events and discrete contingencies, reducing the additional flexible generation or reserve capacity required for reliable operation at high renewable penetration.}

These findings provide practical insights for LDES planning and operation. The full-year analysis supports LDES technology selection under practical design constraints by clarifying the trade-off between RTE and storage duration and the increasing importance of power capacity at longer durations. The \edit{contingency analysis} further \edit{demonstrates} how the seasonal LDES energy state affects corrective operation and subsequent energy recovery under $N-1$ contingencies. Together, the two analyses characterize the role of LDES across both full-year operation and contingency conditions in highly renewable transmission networks.

%%\appendix
%%\section{Example Appendix Section}
%%\label{app1}

%%Appendix text.

%%       \citet{<label>} ==> Lamport (1994)
%%       \citep{<label>} ==> (Lamport, 1994)
%%
%%Example citation, See 



\newpage
\bibliographystyle{elsarticle-harv}
\bibliography{ref}

@article{denholm2021challenges,
  title={The challenges of achieving a 100\% renewable electricity system in the United States},
  author={Denholm, Paul and Arent, Douglas J and Baldwin, Samuel F and Bilello, Daniel E and Brinkman, Gregory L and Cochran, Jaquelin M and Cole, Wesley J and Frew, Bethany and Gevorgian, Vahan and Heeter, Jenny and others},
  journal={Joule},
  volume={5},
  number={6},
  pages={1331--1352},
  year={2021},
  publisher={Elsevier}
}

@article{raynaud2018energy,
  title={Energy droughts from variable renewable energy sources in European climates},
  author={Raynaud, Damien and Hingray, Benoit and Fran{\c{c}}ois, Baptiste and Creutin, Jean Dominique},
  journal={Renewable Energy},
  volume={125},
  pages={578--589},
  year={2018},
  publisher={Elsevier}
}

@article{van2024temporally,
  title={Temporally compounding energy droughts in European electricity systems with hydropower},
  author={van der Most, Lieke and van der Wiel, Karin and Benders, RMJ and Gerbens-Leenes, PW and Bintanja, Richard},
  journal={Nature Energy},
  volume={9},
  number={12},
  pages={1474--1484},
  year={2024},
  publisher={Nature Publishing Group UK London}
}

@article{li2024unseen,
  title={The unseen AI disruptions for power grids: LLM-induced transients},
  author={Li, Yuzhuo and Mughees, Mariam and Chen, Yize and Li, Yunwei Ryan},
  journal={arXiv preprint arXiv:2409.11416},
  year={2024}
}

@inproceedings{lin2024exploding,
  title={Exploding ai power use: an opportunity to rethink grid planning and management},
  author={Lin, Liuzixuan and Wijayawardana, Rajini and Rao, Varsha and Nguyen, Hai and GNIBGA, Emmanuel Wedan and Chien, Andrew A},
  booktitle={Proceedings of the 15th ACM International Conference on Future and Sustainable Energy Systems},
  pages={434--441},
  year={2024}
}

@article{jenkins2017enhanced,
  title={Enhanced decision support for a changing electricity landscape: the GenX configurable electricity resource capacity expansion model},
  author={Jenkins, Jesse D and Sepulveda, Nestor A},
  year={2017},
  publisher={MIT Energy Initiative}
}

@article{sepulveda2021design,
  title={The design space for long-duration energy storage in decarbonized power systems},
  author={Sepulveda, Nestor A and Jenkins, Jesse D and Edington, Aurora and Mallapragada, Dharik S and Lester, Richard K},
  journal={Nature Energy},
  volume={6},
  number={5},
  pages={506--516},
  year={2021},
  publisher={Nature Publishing Group UK London}
}

@article{zhang2020benefit,
  title={Benefit analysis of long-duration energy storage in power systems with high renewable energy shares},
  author={Zhang, Jiazi and Guerra, Omar J and Eichman, Joshua and Pellow, Matthew A},
  journal={Frontiers in Energy Research},
  volume={8},
  pages={527910},
  year={2020},
  publisher={Frontiers Media SA}
}

@article{chu2025long,
  title={Long-duration energy storage in transmission-constrained variable renewable energy systems},
  author={Chu, Andrew K and Baik, Ejeong and Benson, Sally M},
  journal={Cell Reports Sustainability},
  volume={2},
  number={1},
  year={2025},
  publisher={Elsevier}
}

@article{piansky2024long,
  title={Long duration battery sizing, siting, and operation under wildfire risk using progressive hedging},
  author={Piansky, Ryan and Stinchfield, Georgia and Kody, Alyssa and Molzahn, Daniel K and Watson, Jean-Paul},
  journal={Electric Power Systems Research},
  volume={235},
  pages={110785},
  year={2024},
  publisher={Elsevier}
}

@article{gayme2012optimal,
  title={Optimal power flow with large-scale storage integration},
  author={Gayme, Dennice and Topcu, Ufuk},
  journal={IEEE Transactions on Power Systems},
  volume={28},
  number={2},
  pages={709--717},
  year={2012},
  publisher={IEEE}
}

@article{gabash2012active,
  title={Active-reactive optimal power flow in distribution networks with embedded generation and battery storage},
  author={Gabash, Aouss and Li, Pu},
  journal={IEEE Transactions on Power Systems},
  volume={27},
  number={4},
  pages={2026--2035},
  year={2012},
  publisher={IEEE}
}

@inproceedings{xu2001optimal,
  title={Optimal load shedding strategy in power systems with distributed generation},
  author={Xu, Ding and Girgis, Adly A},
  booktitle={2001 IEEE Power Engineering Society Winter Meeting. Conference Proceedings (Cat. No. 01CH37194)},
  volume={2},
  pages={788--793},
  year={2001},
  organization={IEEE}
}

@article{hazra2007congestion,
  title={Congestion management using multiobjective particle swarm optimization},
  author={Hazra, Jagabondhu and Sinha, Avinash K},
  journal={IEEE Transactions on Power Systems},
  volume={22},
  number={4},
  pages={1726--1734},
  year={2007},
  publisher={IEEE}
}

@article{majumdar1996interruptible,
  title={Interruptible load management using optimal power flow analysis},
  author={Majumdar, S and Chattopadhyay, D and Parikh, Jyoti},
  journal={IEEE Transactions on Power Systems},
  volume={11},
  number={2},
  pages={715--720},
  year={1996},
  publisher={IEEE}
}

@article{martins2008redispatch,
  title={Redispatch to reduce rotor shaft impacts upon transmission loop closure},
  author={Martins, Nelson and de Oliveira, Edimar Jose and Moreira, Weberson Carvalho and Pereira, Jos{\'E} Luiz Rezende and Fontoura, Rafael Montes},
  journal={IEEE Transactions on Power Systems},
  volume={23},
  number={2},
  pages={592--600},
  year={2008},
  publisher={IEEE}
}

@article{thukaram2008optimal,
  title={Optimal reactive power dispatch in a large power system with AC--DC and FACTS controllers},
  author={Thukaram, D and Yesuratnam, G},
  journal={IET generation, transmission \& distribution},
  volume={2},
  number={1},
  pages={71--81},
  year={2008},
  publisher={IET}
}

@article{stott2005security,
  title={Security analysis and optimization},
  author={Stott, Brian and Alsac, Ongun and Monticelli, Alcir J},
  journal={Proceedings of the IEEE},
  volume={75},
  number={12},
  pages={1623--1644},
  year={2005},
  publisher={IEEE}
}

@article{capitanescu2011state,
  title={State-of-the-art, challenges, and future trends in security constrained optimal power flow},
  author={Capitanescu, Florin and Ramos, JL Martinez and Panciatici, Patrick and Kirschen, Daniel and Marcolini, A Marano and Platbrood, Ludovic and Wehenkel, Louis},
  journal={Electric power systems research},
  volume={81},
  number={8},
  pages={1731--1741},
  year={2011},
  publisher={Elsevier}
}

@inproceedings{marley2016multi,
  title={Multi-period AC-QP optimal power flow including storage},
  author={Marley, Jennifer F and Hiskens, Ian A},
  booktitle={2016 Power Systems Computation Conference (PSCC)},
  pages={1--7},
  year={2016},
  organization={IEEE}
}

@article{soares2017active,
  title={Active distribution grid management based on robust AC optimal power flow},
  author={Soares, Tiago and Bessa, Ricardo J and Pinson, Pierre and Morais, Hugo},
  journal={IEEE Transactions on Smart Grid},
  volume={9},
  number={6},
  pages={6229--6241},
  year={2017},
  publisher={IEEE}
}

@article{alizadeh2022envisioning,
  title={Envisioning security control in renewable dominated power systems through stochastic multi-period AC security constrained optimal power flow},
  author={Alizadeh, Mohammad Iman and Usman, Muhammad and Capitanescu, Florin},
  journal={International Journal of Electrical Power \& Energy Systems},
  volume={139},
  pages={107992},
  year={2022},
  publisher={Elsevier}
}

@misc{cpuc,
  author       = {{California Public Utilities Commission}},
  title        = {{Decision Granting, with Modifications, Long Duration Energy Storage Council's Petition for Modification 21-06-035}},
  howpublished = {{California Public Utilities Commission}},
  year         = {2025},
  note         = {Accessed: May 26, 2026}
}

@book{nelsen2006introduction,
  title={An introduction to copulas},
  author={Nelsen, Roger B},
  year={2006},
  publisher={Springer}
}

@article{momoh2002economic,
  title={Economic operation and planning of multi-area interconnected power systems},
  author={Momoh, JA and Dias, LG and Guo, SX and Adapa, R},
  journal={IEEE transactions on power systems},
  volume={10},
  number={2},
  pages={1044--1053},
  year={2002},
  publisher={IEEE}
}

@article{ge1999optimal,
  title={Optimal active power flow incorporating power flow control needs in flexible AC transmission systems},
  author={Ge, SY and Chung, TS},
  journal={IEEE Transactions on Power Systems},
  volume={14},
  number={2},
  pages={738--744},
  year={1999},
  publisher={IEEE}
}

@inproceedings{geth2020flexible,
  title={A flexible storage model for power network optimization},
  author={Geth, Frederik and Coffrin, Carleton and Fobes, David},
  booktitle={Proceedings of the Eleventh ACM International Conference on future energy systems},
  pages={503--508},
  year={2020}
}

@article{seabold2010statsmodels,
  title={Statsmodels: econometric and statistical modeling with python.},
  author={Seabold, Skipper and Perktold, Josef and others},
  journal={scipy},
  volume={7},
  number={1},
  pages={92--96},
  year={2010}
}

@article{sparks2018nasapower,
  title={nasapower: a NASA POWER global meteorology, surface solar energy and climatology data client for R},
  author={Sparks, Adam H},
  journal={Journal of Open Source Software},
  volume={3},
  number={30},
  pages={1035},
  year={2018}
}

@techreport{wan2010development,
  title={Development of an equivalent wind plant power-curve},
  author={Wan, Yih-Huei and Ela, Erik and Orwig, Kirsten},
  year={2010},
  institution={National Renewable Energy Laboratory (NREL), Golden, CO (United States)}
}

@article{kim2023revealing,
  title={Revealing the impact of renewable uncertainty on grid-assisted power-to-X: A data-driven reliability-based design optimization approach},
  author={Kim, Jeongdong and Qi, Meng and Park, Jinwoo and Moon, Il},
  journal={Applied Energy},
  volume={339},
  pages={121015},
  year={2023},
  publisher={Elsevier}
}

@article{birchfield2016grid,
  title={Grid structural characteristics as validation criteria for synthetic networks},
  author={Birchfield, Adam B and Xu, Ti and Gegner, Kathleen M and Shetye, Komal S and Overbye, Thomas J},
  journal={IEEE Transactions on power systems},
  volume={32},
  number={4},
  pages={3258--3265},
  year={2016},
  publisher={IEEE}
}

@inproceedings{li2018load,
  title={Load modeling in synthetic electric grids},
  author={Li, Hanyue and Bornsheuer, Ashly L and Xu, Ti and Birchfield, Adam B and Overbye, Thomas J},
  booktitle={2018 IEEE Texas Power and Energy Conference (TPEC)},
  pages={1--6},
  year={2018},
  organization={IEEE}
}

@article{johnson2025examodelspower,
  title={ExaModelsPower. jl: A GPU-Compatible Modeling Library for Nonlinear Power System Optimization},
  author={Johnson, Sanjay and Lauinger, Dirk and Shin, Sungho and Pacaud, Fran{\c{c}}ois},
  journal={arXiv preprint arXiv:2510.12897},
  year={2025}
}

@article{shin2024accelerating,
  title={Accelerating optimal power flow with GPUs: SIMD abstraction of nonlinear programs and condensed-space interior-point methods},
  author={Shin, Sungho and Anitescu, Mihai and Pacaud, Fran{\c{c}}ois},
  journal={Electric Power Systems Research},
  volume={236},
  pages={110651},
  year={2024},
  publisher={Elsevier}
}

@manual{caisoBPMFNM,
  title        = {Business Practice Manual for Managing Full Network Model},
  organization = {California Independent System Operator},
  address      = {Folsom, CA},
  year         = {2026},
  url          = {https://bpmcm.caiso.com/pages/default.aspx},
  note         = {Last revised: 2026}
}

@inproceedings{huang2002voltage,
  title={Voltage stability constrained load curtailment procedure to evaluate power system reliability measures},
  author={Huang, Garng M and Nair, N-KC},
  booktitle={2002 IEEE Power Engineering Society Winter Meeting. Conference Proceedings (Cat. No. 02CH37309)},
  volume={2},
  pages={761--765},
  year={2002},
  organization={IEEE}
}

@techreport{nerc2016reactive,
  author       = {{North American Electric Reliability Corporation}},
  title        = {Reliability Guideline: Reactive Power Planning},
  institution  = {North American Electric Reliability Corporation (NERC)},
  year         = {2016},
  month        = dec,
  url          = {https://www.nerc.com/globalassets/who-we-are/standing-committees/rstc/sams/reliability-guideline---reactive-power-planning.pdf}
}

@article{federal2016reactive,
  title={Reactive power requirements for non-synchronous generation},
  author={Federal Energy Regulatory Commission and others},
  journal={Docket No. RM16-1-000, Order No},
  volume={827},
  year={2016}
}

@techreport{karlson2012wind,
  title={Wind power plant short-circuit modeling guide.},
  author={Karlson, Benjamin and Williams, Joseph},
  year={2012},
  institution={Sandia National Laboratories}
}

@article{viswanathan20222022,
  title={2022 grid energy storage technology cost and performance assessment},
  author={Viswanathan, Vilayanur and Mongird, Kendall and Franks, Ryan and Li, Xiaolin and Sprenkle, Vincent and Baxter, Richard},
  journal={Energy},
  volume={2022},
  pages={1--151},
  year={2022}
}

@inproceedings{baker2021never,
  title={Solutions of {DC OPF} are Never {AC} Feasible},
  author={Baker, Kyri},
  booktitle={Proceedings of the Twelfth ACM International Conference on Future Energy Systems},
  pages={264--268},
  year={2021},
  publisher={Association for Computing Machinery}
}

@article{dowling2020role,
  title={Role of Long-Duration Energy Storage in Variable Renewable Electricity Systems},
  author={Dowling, Jacqueline A and Rinaldi, Katherine Z and Ruggles, Tyler H and Davis, Steven J and Yuan, Mengyao and Tong, Fan and Lewis, Nathan S and Caldeira, Ken},
  journal={Joule},
  volume={4},
  number={9},
  pages={1907--1928},
  year={2020}
}

@article{kittel2026longduration,
  title={Long-duration electricity storage needs for coping with {Dunkelflaute} events in {Europe}},
  author={Kittel, Martin and Roth, Alexander and Schill, Wolf-Peter},
  journal={Nature Communications},
  volume={17},
  pages={4210},
  year={2026}
}

@book{billinton1996reliability,
  title={Reliability Evaluation of Power Systems},
  author={Billinton, Roy and Allan, Ronald N},
  edition={2},
  year={1996},
  publisher={Plenum Press},
  address={New York}
}

@techreport{iea2021netzero,
  title={Net Zero by 2050: A Roadmap for the Global Energy Sector},
  author={{International Energy Agency}},
  institution={International Energy Agency},
  address={Paris},
  year={2021}
}
\end{document}